\documentclass{article}

\usepackage{PRIMEarxiv}

\usepackage[utf8]{inputenc} 
\usepackage[T1]{fontenc}    
\usepackage{url}            
\usepackage{booktabs}       
\usepackage{amsfonts}       
\usepackage{nicefrac}       
\usepackage{microtype}      
\usepackage{lipsum}
\usepackage{fancyhdr}       
\usepackage{graphicx}       
\graphicspath{{media/}}     

\usepackage[ruled,vlined,linesnumbered,noend]{algorithm2e}    
\usepackage{amsfonts}
\usepackage{amsmath}
\usepackage{amssymb}
\usepackage{amsthm}
\usepackage{boxedminipage}
\usepackage{caption} 
\usepackage{float}

\usepackage{enumitem}
\usepackage{graphicx}
\usepackage{multirow}
\usepackage{natbib}  
\usepackage{pifont}
\usepackage{pgfplots}
\usepackage{comment}
\usepackage{xcolor}         
\usepackage{tikz}
\usepackage{thm-restate}
\setlist{nolistsep}
\usetikzlibrary{arrows}
\usetikzlibrary{patterns}
\usepackage[hidelinks]{hyperref}
\usepackage{cleveref}

\newtheorem{lemma}{Lemma}%
\newtheorem{definition}{Definition}%
\newtheorem{proposition}{Proposition}%
\newtheorem{example}{Example}

\usepackage[color={red!100!green!33},textsize=scriptsize]{todonotes}

\newcommand{\Ch}{\mathrm{Ch}}

\pgfplotsset{compat=1.18}

\allowdisplaybreaks

\usepackage{authblk}

\title{\bf Probabilistic Analysis of Stable Matching in
\\ Large Markets with Siblings}

\author[1,2]{Zhaohong Sun}
\author[3]{Tomohiko Yokoyama}
\author[1]{Makoto Yokoo}

\affil[1]{Kyushu University, Japan}
\affil[2]{CyberAgent, Japan}
\affil[3]{The University of Tokyo, Japan}

\date{\vspace{-10mm}}

\begin{document}

\maketitle
\begin{abstract}
We study a practical centralized matching problem which assigns children to daycare centers.
The collective preferences of siblings from the same family introduce complementarities, which can lead to the absence of stable matchings, as observed in the hospital-doctor matching problems involving couples. Intriguingly, stable matchings are consistently observed in real-world daycare markets, despite the prevalence of sibling applicants.

We conduct a probabilistic analysis of large random markets to examine the existence of stable matchings in such markets. Specifically, we examine scenarios where daycare centers have similar priorities over children, a common characteristic in real-world markets. Our analysis reveals that as the market size approaches infinity, the likelihood of stable matchings existing converges to 1.

To facilitate our exploration, we refine an existing heuristic algorithm to address a more rigorous stability concept, as the original one may fail to meet this criterion. Through extensive experiments on both real-world and synthetic datasets, we demonstrate the effectiveness of our revised algorithm in identifying stable matchings, particularly when daycare priorities exhibit high similarity.
\end{abstract}

\section{Introduction}
Stability is a foundational concept in preference-based matching theory \citep{RoSo90a}, with significant implications for both theoretical frameworks and practical applications \citep{Roth08a}. Its importance was underscored by the awarding of the 2012 Nobel Prize in Economics. This fundamental concept is crucial for the success of various markets, including the National Resident Matching Program \citep{Roth84a} and public school choice programs \citep{AbSo03b, APRS05a}.

Despite its significance, the challenge posed by complementarities in preferences often leads to the absence of a stable matching. A persistent issue in this context is the incorporation of couples into centralized clearing algorithms for professionals like doctors and psychologists \citep{RoPe99a}. Couples typically view pairs of jobs as complements, which can result in the non-existence of a stable matching \citep{Roth84a, KlKl05a}. Moreover, verifying the existence of a stable matching is known to be NP-hard, even in restrictive settings \citep{Ronn90a, McMa10a, BMM14a}.

Nevertheless, real-life markets of substantial scale do exhibit stable matchings even in the presence of couples. For example, in the psychologists' markets, couples constituted only about $1\%$ of all participants from 1999 to 2007 \citep{KPR13a}. \citet{ABH14a} demonstrate that if the proportion of couples grows sufficiently slowly compared to the number of single doctors, then a stable matching is very likely to exist in a large market.

In this paper, we shift our attention to daycare matching markets in Japan, where the issue of waiting children has become one of the most urgent social challenges due to the scarcity of daycare facilities \citep{KaKo23a}. The daycare matching problem is a natural extension of matching with couples, with the notable distinction that the number of siblings in each family can exceed two. We are actively collaborating with multiple municipalities, providing advice to design and implement new centralized algorithms tailored to their specific needs.

The objective of this research is to gain a more nuanced understanding of why stable matchings exist in practical daycare markets. Recently, stable matchings have been reported in these markets where optimization approaches are utilized \citep{STM+23a, SYT+24a}, but the underlying reasons have not been thoroughly examined.
Furthermore, theoretical guarantees established in prior research on matching with couples may not readily extend to the daycare market \citep{KPR13a,ABH14a}, primarily due to two key factors. Firstly, a distinctive characteristic of Japanese daycare markets is the substantial proportion, approximately $20\%$, of children with siblings. This stands in contrast to the assumption of near-linear growth of couples in previous research. Secondly, we consider a stronger stability concept than the previous one,  tailored for daycare markets. Our proposal has been presented to government officials and esteemed economists, who concur that this modification better suits the daycare markets.

Our contributions can be summarized as follows:

Firstly, we propose an Extended Sorted Deferred Acceptance (ESDA) algorithm, which builds upon the existing heuristic Sorted Deferred Acceptance (SDA) algorithm \citep{ABH14a}. The modification is necessary because the original algorithm may fail to produce a matching that satisfies our stricter stability concept (Theorem~\ref{theo:SDA:original}).  We further demonstrate that the ESDA algorithm yields a stable matching when it terminates successfully (Theorem~\ref{theo:ESDA:stability}). 

Second, we conduct a probabilistic analysis to investigate the existence of stable matchings in large random daycare markets, modeled using probability distributions. A key observation is that, in practice, daycares often share similar priority structures over children. Our main result demonstrates that in such random markets, the probability of a stable matching existing approaches 1 as the market size becomes infinitely large (Theorem~\ref{theorem:main}). To the best of our knowledge, this is the first study to provide a theoretical framework that explains the consistent presence of stable matchings in real-world daycare markets.

Third, we conduct comprehensive experiments on both real-world datasets and a diverse range of synthetic datasets.  
Our results demonstrate that a stable matching is highly likely to exist, and the ESDA algorithm remains highly effective, particularly in scenarios where daycare priorities exhibit significant similarity.

\section{Related Work}

\cite{Ronn90a} initially established that verifying the existence of stable matchings in the presence of couples is an NP-hard problem, even if each hospital offers only one position. Follow-up work by \cite{McMa10a} showed this computational intractability result still holds even when couples' preferences are limited to pairs of positions within the same hospital. Furthermore, \cite{BIS11a} demonstrated that it remains NP-hard
when all doctors are ranked according to a common order adopted by all hospitals.

A classical work on matching with couples, conducted by \cite{KPR13a}, illustrates that as the market size approaches infinity, the probability of a stable matching existing converges to 1, given the growth rate of couples is suitably slow in relation to the market size, e.g., when the number of couples is $\sqrt{n}$ where $n$ represents the number of singles. \cite{ABH14a} propose an improved matching algorithm, building on the foundation laid by \cite{KPR13a}. This refined algorithm demonstrates that, even if the number of couples grows at a near-linear rate of $n^{\epsilon}$ with $0 < \epsilon < 1$, a stable matching can still be found with high probability. In contrast, \cite{ABH14a} highlight that as the number of couples increases at a linear rate, the probability of a stable matching existing diminishes significantly.

\cite{KPR13a} devised the Sequential Couples Algorithm to address matching problems involving couples, which follows a three-step procedure. First, it computes a stable matching without considering couples, using the DA algorithm. Next, it handles each couple according to a predefined order denoted as $\pi$. Single doctors displaced by couples are accommodated one by one, allowing them to apply to hospitals based on their preferences. However, if an application is made to a hospital where any member of a couple has previously submitted an application, the algorithm declares a unsuccessful termination, even though a stable matching may indeed exist.

The Sorted Deferred Acceptance (SDA) algorithm, as introduced by~\cite{ABH14a}, follows a similar trajectory to the Sequential Couples Algorithm. 
We extend its application to the context of daycare matching with siblings. The algorithm begins by computing a stable matching without considering families with siblings, denoted as $F^S$, using the DA algorithm. Subsequently, it sequentially processes each family, denoted as $f$, based on a predefined order denoted as $\pi$. Children without siblings who are displaced by family $f$ are processed individually, enabling them to apply to daycare centers according to their preferences. If any child from family $f'\in F^S$ with siblings is affected during this process, a new order $\pi'$ is attempted, with $f$ being inserted before $f'$. The algorithm terminates and returns an unsuccessful termination if any child from family $f$ is affected or if the same permutation has been attempted twice.

One potential solution to overcome the non-existence of stable matchings is to explore restricted preference domains. In this regard, \cite{KlKl05a} investigated a restricted preference domain known as weak responsiveness, ensuring the presence of stable matchings in the presence of couples. \cite{HaKo10a} introduced the concept of ``bilateral substitute'' within the framework of matching with contracts \cite{HaMi05a}, encompassing matching with couples as a specific case, and they demonstrated that weak responsiveness implies bilateral substitutes.

In practical applications, the National Resident Matching Program employed a heuristic based on the incremental algorithm proposed by \cite{RoVa90a}. \cite{BFI16a} proposed a different approach  involves the utilization of the Scarf algorithm \cite{Scar67a} to identify a fractional matching. If the outcome proves to be integral, it is then considered a stable matching. Moreover, researchers have explored the application of both integer programming and constraint programming to address the complexities of matching with couples \cite{MOPU07a,BMM14a,MMT17a}. Notably, these methodologies have recently been adapted in the daycare matching market as well \cite{STM+23a,SYT+24a}.

Another trend in the literature explores the combination of bandit algorithms with matching market design. In these studies, preferences are initially unknown and are learned through the interactions between the two sides of agents (see \cite{DaKa05a,LMJ20a,BSS21a,LRMJ21a,JWWJ+21a,KYL22a}). This contrasts with our setting, where preferences and priorities are submitted to the system in advance.

\section{Preliminaries}

In this section, we present the framework of a daycare market, expanding upon the classical problem of hospital-doctor matching with couples. We also generalize three fundamental properties that have been extensively examined in the literature of two-sided matching markets.

\subsection{Model}
\label{sec:model}
The daycare matching problem is represented by the tuple $I = (C, F, D, Q, \succ_F, \succ_D)$, where $C$, $F$ and $D$ denote sets of children, families, and daycare centers, respectively.

Each child $c \in C$ belongs to a family denoted as $f_c \in F$. Each family $f \in F$ is associated with a subset of children, denoted as $C_f \subseteq C$. 
In cases where a family contains more than one child, e.g., $C_f = (c_1, c_2, \ldots, c_{k})$ with $k > 1$, these siblings are arranged in a predefined order, such as by age.


Let $D$ represent a set of daycare centers, referred to as ``daycares'' for brevity. A dummy daycare denoted as $d_0$ is included in $D$, signifying the possibility of a child being unmatched. Each individual daycare $d$ establishes a quota, denoted as $Q_d$, where the symbol $Q$ represents all quotas.


Each family $f$ reports a strict \emph{preference ordering} $\succ_f$, defined over tuples of daycare centers, reflecting the collective preferences of the children within $C_f$. The notation $D(\succ_f, j)$ is used to represent the $j$-th tuple of daycares in $\succ_f$, and the overall preference profile of all families is denoted as $\succ_F$.
\begin{example}
Consider family $f$ with $C_f = (c_1, c_2, \ldots, c_{k})$ where the children are arranged in a predetermined order. A tuple of daycares in $\succ_f$, denoted as $(d_1^*, d_2^*, \ldots, d^*_{k})$, indicates that for each $i \in \{1, 2, \ldots, k\}$, child $c_i$ is matched to some daycare $d_i^* \in D$. It's possible that $d_i^* = d_j^*$, indicating that both child $c_i$ and child $c_j$ are matched to daycare $d_i^*$.
\end{example}


Each daycare $d \in D$ maintains a strict \emph{priority ordering} $\succ_d$ over $C \cup \{\emptyset\}$, encompassing both the set of children $C$ and an empty option. A child $c \in C$ is considered acceptable to daycare $d$ if $c \succ_d \emptyset$, and deemed unacceptable if $\emptyset \succ_d c$. The priority profile of all daycares is denoted as $\succ_D$.

%




A \emph{matching} is defined as a function $\mu: C \cup D \rightarrow C \cup D$ such that i) $\forall c\in C$, $\mu(c) \in D$, ii) $\forall d\in D$, $\mu(d) \subseteq C$ and iii) $\mu(c) =d$ if and only if $c\in \mu(d)$. For a given matching $\mu$, the assignment of a child $c$ is denoted by $\mu(c)$, and the set of children assigned to a daycare $d$ is denoted by $\mu(d)$. For a family $f$ with children $C_f = (c_1, c_2, \ldots, c_k)$, the family’s assignment is represented as $\mu(f) = \bigl(\mu(c_1), \mu(c_2), \ldots, \mu(c_k)\bigr)$.

\subsection{Fundamental Properties}


The first property, individual rationality, stipulates that each family is matched to some tuple of daycares that are weakly better than being unmatched, and no daycare is matched with an unacceptable child. It is noteworthy that each family is considered an agent, rather than individual children.

\begin{definition}[Individual Rationality]
A \emph{matching} $\mu$ satisfies individual rationality if two conditions hold: i) $\forall f\in F, \mu(f) \succ_f (d_0, d_0,\ldots, d_0)$ or $\mu(f) = (d_0, d_0, \ldots, d_0)$, and ii) $\forall d\in D, \forall c \in \mu(d), c \succ_d \emptyset$.
\end{definition}

Feasibility in Definition~\ref{def:feasibility} necessitates that i) each child is assigned to one daycare including the dummy daycare $d_0$, and ii) the number of children matched to each daycare $d$ does not exceed its specific quota $Q_d$.

\begin{definition}[Feasibility]
\label{def:feasibility}
A matching $\mu$ is \emph{feasible} if it satisfies the following conditions: i) $\forall c\in C$, $\lvert \mu(c) \rvert = 1$, and ii) $\forall d \in D$,
$\lvert\mu(d) \rvert\leq Q_d$.
\end{definition}

Stability is a well-explored solution concept within the domain of two-sided matching theory. Before delving into its definition, we introduce the concept of a \emph{choice function} as outlined in Definition~\ref{def:choice_func}. It captures the intricate process by which daycares select children, capable of incorporating various considerations such as priority, diversity goals, and distributional constraints (see, e.g., \citep{HaMi05a,AzSu21a,STY+23a,KaKo23a}). Following the work by \citep{ABH14a},
our choice function operates through a greedy selection of children based on priority only, simplifying the representation of stability.

\begin{definition}[Choice Function of a Daycare]
\label{def:choice_func}
    For a given set of children $C' \subseteq C$, the \emph{choice function} of daycare $d$, denoted as $\mathrm{Ch}_d: C' \rightarrow 2^{C'}$, selects children one by one in descending order of $\succ_d$ without exceeding quota $Q_d$.
\end{definition}

In this paper, we explore a slightly stronger stability concept than the original one studied in \citep{ABH14a}. It extends the idea of eliminating blocking pairs~\citep{GaSh62a} to address the removal of blocking coalitions between families and a selected subset of daycares. 

\begin{definition}[Stability]\label{def:stability}
Given a feasible and individually rational matching $\mu$, family $f$ with children $C_f = (c_1,c_2,\ldots,c_{k})$ and the $j$-th tuple of daycares $D(\succ_f, j)=
    (d_1^*, d_2^*, \ldots, d_{k}^*)$ in $\succ_f$, form a \emph{blocking coalition} if the following two conditions hold, 
    \begin{description}
        \item [1.]family $f$ prefers $(d_1^*$, $d_2^*$, $\ldots$, $d_{k}^*)$  to its current assignment $\mu(f)$, i.e., $D(\succ_f, j)$  $\succ_f \mu(f)$, and\\
        \item [2.]for each distinct daycare $d$ from $(d_1^*,d_2^*,\ldots,d_{k}^*)$, we have $C(\succ_f, j, d)$ $\subseteq$ $\mathrm{Ch}_d( (\mu(d) \setminus C_f) \cup C(\succ_f, j, d))$, where $C(\succ_f, j, d) \subseteq C_f$ denotes a subset of children from family $f$ who apply to daycare $d$ with respect to $D(\succ_f, j)$.
    \end{description}
 %
A feasible and individually rational matching satisfies stability if no blocking coalition exists.
\end{definition}

Consider the input to $\mathrm{Ch}_d(\cdot)$ in Condition 2 in Definition~\ref{def:stability}. First, we calculate $\mu(d) \setminus C_f$, representing the children matched to $d$ in matching $\mu$ but not from family $f$. Then, we consider $C(\succ_f, j, d)$, which denotes the subset of children from family $f$ who apply to $d$ according to the tuple of daycares $D(\succ_f, j)$.

This process accounts for situations where a child $c$ is paired with $d$ in $\mu$ but is not included in $C(\succ_f, j, d)$, indicating that $c$ is applying to a different daycare $d'\neq d$ according to $D(\succ_f, j)$. Consequently, child $c$ has the flexibility to pass his assigned seat from $d$ to his siblings in need. Otherwise, child $c$ would compete with his siblings for seats at $d$ despite he intends to apply elsewhere. 

\subsection{Motivation of New Stability Concept}
\label{appendix:motivation_stability}

The primary reason for modifying the stability concept lies in the differing selection criteria between hospital-doctor matching and daycare allocation. In the hospital-doctor matching problem, hospitals have preferences over doctors. In contrast, daycare centers use priority orderings based on priority scores to determine which child should be given higher precedence. The priority scoring system is designed to eliminate justified envy and achieve a fair outcome, treating daycare slots as resources to be allocated equitably.
Additionally, it is crucial that siblings do not envy each other, especially when they are not enrolled in the same daycare. Allowing children to transfer their seats to other siblings can potentially reduce waste and increase overall welfare.
We presented this new stability concept to multiple government officials from different municipalities and several renowned economists. They all agreed that the modification is more appropriate for the daycare matching setting.

On the other hand, the stability concept by \citep{ABH14a} does not take siblings' assignments into account.  To distinguish it from our concept, we refer to their stability as ABH-stability, named after the authors' initials. 
The ABH-stability concept was originally designed for matching with couples and defined by enumerating all possible scenarios.
In Definition~\ref{def:ABH-stability}, we consolidate these scenarios into a concise format, which highlights the differences from our definition. The primary distinction from Definition~\ref{def:stability} lies in the input to $\Ch_d(\cdot)$ in Condition 2: it uses $\mathrm{Ch}_d(\mu(d) \cup C(\succ_f, j, d))$, instead of $\mathrm{Ch}_d(\mu(d) \setminus C_f \cup C(\succ_f, j, d))$.

\begin{definition}[ABH-Stability]
\label{def:ABH-stability}
Given a feasible and individually rational matching $\mu$, family $f$ with children $C_f = (c_1,c_2,\ldots,c_{k})$ and the $j$-th tuple of daycares $D(\succ_f, j)$ $=$
$(d_1^*,d_2^*,\ldots,d_{k}^*)$ in $\succ_f$, form a \emph{ABH blocking coalition} if the following two conditions hold, 
\begin{description}
    \item[1.] family $f$ prefers $(d_1^*$, $d_2^*$, $\ldots$, $d_{k}^*)$  to its current assignment $\mu(f)$, i.e., $D(\succ_f, j)$  $\succ_f \mu(f)$, and\\
    \item[2.] for each distinct daycare $d$ from  $(d_1^*,d_2^*,\ldots,d_{k}^*)$, we have
 $C(\succ_f, j, d)$ $\subseteq$ $\mathrm{Ch}_d( \mu(d) \cup C(\succ_f, j, d))$,  
 where $C(\succ_f, j, d) \subseteq C_f$ denotes a subset of children from family $f$ who apply to daycare $d$ with respect to $D(\succ_f, j)$.
\end{description}
%
A feasible and individually rational matching satisfies ABH-stability if no ABH blocking coalition exists.
\end{definition}


\begin{proposition}
\label{prop:connection:stability}
    Stability implies ABH-stability, but not vice versa.
\end{proposition}

\begin{proof}
For an ABH blocking coalition, Condition 2 specifies that when a daycare $d$ selects from $\mu(d) \cup C(\succ_f, j, d)$, which includes both its currently matched children and new applicants from family $f$, all children in $C(\succ_f, j, d)$ can be chosen. In contrast, for a blocking coalition as defined in Definition~\ref{def:stability}, daycare $d$ chooses from applicants in $(\mu(d) \setminus C_f) \cup C(\succ_f, j, d)$, which excludes children from family $f$ currently matched to $d$, along with new applicants from $Y_f'$. In simpler terms, if some siblings from family $f$ do not apply to daycare $d$, their seats are freed up, reducing competition for the siblings who do apply. Therefore, if an ABH blocking coalition exists, it also constitutes a blocking coalition under our definition. Conversely, if no blocking coalition exists, there can be no ABH blocking coalition, as shown in Example~\ref{example:stable}. This demonstrates that stability, as defined in Definition~\ref{def:stability}, implies ABH-stability.
\end{proof}

We next illustrate the differences between stability in Definition~\ref{def:stability} and ABH-stability in Definition~\ref{def:ABH-stability} through 
Example~\ref{example:stable}. 

\begin{example}[Comparison of Two Stability Concepts]
    \label{example:stable}
Consider a family $f$ with two children $C_f = (c_1, c_2)$ and three daycare centers $D = \{d_0, d_1, d_2\}$. The daycares $d_1$ and $d_2$ each have one available slot, while the dummy daycare $d_0$ has unlimited capacity. The preferences of family $f$ are $(d_1, d_2) \succ_f (d_2, d_0)$. Each daycare ranks $c_1$ higher than $c_2$.

The matching $(d_2, d_0)$, which assigns $c_1$ to $d_2$ and $c_2$ to $d_0$, is considered ABH-stable by Definition~\ref{def:ABH-stability}. However, it does not satisfy our stricter stability criteria defined in Definition~\ref{def:stability}. This is because it is blocked by family $f$ and the pair $(d_1, d_2)$: child $c_1$ could transfer their seat at $d_2$ to $c_2$, allowing both children to achieve a more preferred assignment.

We consider the matching $(d_1, d_2)$ superior, as it assigns family $f$ to their top choice without negatively impacting any other family. In contrast, the matching $(d_2, d_0)$ results in a wasted seat at daycare $d_1$ and leaves family $f$ unsatisfied.
\end{example}

\citet{KaKo17b} explores two stability concepts, namely strong stability and weak stability, in the context of hospital and doctor matching with distributional constraints. These constraints extend beyond simple capacity limits to include considerations such as regional caps that restrict the number of doctors assigned to specific areas. Both stability concepts accommodate certain forms of blocking pairs. Under strong stability, any blocking pair that does not violate feasibility constraints is deemed valid. In contrast, weak stability introduces a hypothetical scenario in which a blocking doctor is temporarily added to a hospital without being removed from their current assignment, allowing for a broader interpretation of stability.

While the idea of modifying assignment bears some similarity, these two stability concepts differ from ours in two significant ways. First, both strong and weak stability are less stringent than the traditional notion of stability, as they allow certain types of blocking pairs to persist. In contrast, our stability concept is stricter than the original, as it eliminates even weaker forms of blocking coalitions. Second, our approach accounts for scenarios where children can transfer their seats to their siblings without negatively impacting other families. In comparison, their framework focuses on whether an individual agent can be reassigned to a more favorable option without violating feasibility constraints.

\subsection{Non-existence of Stable Matchings in Theory}
It is well-known that a stable matching is not guaranteed when couples exist \citep{Roth84a}. We provide an example to illustrate that even when each family has at most two children, and all daycares have the same priority ordering over children, a stable matching may not exist. 


\begin{example}[Non-existence of Stable Matchings]
    \label{example:non_existence_stable}
Consider three families: $f_1$ with children $C_{f_1} = (c_1, c_2)$, $f_2$ with children $C_{f_2} = (c_3, c_4)$,
and $f_3$ with children $C_{f_3} = (c_5, c_6)$.
There are four daycares: $D = \{d_0, d_1, d_2, d_3\}$, each with a single slot except for a dummy daycare $d_0$. The preference profile of the families and the priority profile of the daycares are as follows:
\begin{align*}
    & \succ_{f_1}: (d_1, d_2) \quad \succ_{f_2}: (d_2, d_3) \quad \succ_{f_3}: (d_3, d_1)
    \\
    & \forall d \in D \ \succ_{d}: c_1,c_6,c_3,c_2,c_5,c_4 
\end{align*}

There are three feasible matchings except for the empty matching which can not be stable, namely: 
\begin{itemize}
    \item Matching $\mu_1$ where $\mu_1(f_1) = (d_1, d_2)$, $\mu_1(f_2) = (d_0, d_0)$, and $\mu_1(f_3) = (d_0, d_0)$.
    \item Matching $\mu_2$ where $\mu_2(f_1) = (d_0, d_0)$, $\mu_2(f_2) = (d_2, d_3)$, and $\mu_2(f_3) = (d_0, d_0)$.
    \item Matching $\mu_3$ where $\mu_3(f_1) = (d_0, d_0)$, $\mu_3(f_2) = (d_0, d_0)$, and $\mu_3(f_3) = (d_3, d_1)$.
\end{itemize}

Matching $\mu_1$ cannot be stable, because family $f_2$ could form a blocking coalition with a pair of daycares $(d_2, d_3)$, where $\Ch_{d_2}(\{c_2, c_3\})$ $= \{c_3\}$ and $\Ch_{d_3}(\{c_4\})$ $= \{c_4\}$. 
Similarly, matching $\mu_2$ is blocked by family $f_3$ and daycares $(d_3, d_1)$, and matching $\mu_3$ is blocked by family $f_1$ and daycares $(d_1, d_2)$.
Consequently, none of the matchings $\mu_1$, $\mu_2$, and $\mu_3$ is stable.
\end{example}

\section{Extended Sorted Deferred Acceptance (ESDA)}\label{section:ESDA}

In this section, we introduce the Extended Sorted Deferred Acceptance (ESDA) algorithm, a heuristic method demonstrated to be effective in computing stable matchings across diverse real-world and synthetic datasets. Importantly, the ESDA algorithm serves as a foundational component in our probability analysis for large random markets.

\subsection{Rejection Chain and Cycle}

The Deferred Acceptance (DA) algorithm is a classical algorithm in matching theory under preferences~\citep{GaSh62a,Roth85a}. The (children-proposing) DA algorithm proceeds iteratively through the following two phases. In the application phase, children first apply to their most preferred daycares that have not rejected them so far. 
In the selection phase, each daycare selects children based on its priorities from the pool of new applicants in the current round and the temporarily matched children from the previous round without exceeding specific quotas.
The algorithm terminates when no child submits any further applications. 
%

We now introduce two concepts, rejection chain and rejection cycle, which play an important role in both the design of our ESDA algorithm and its theoretical analysis.

\begin{definition}[Rejection Chain]
    \label{def:rejection_chain}
When a child $c_1^*$ applies to a daycare $d_1^*$ that is already at full capacity, daycare $d_1^*$ must reject some child $c_2^*$ (which could be $c_1^*$). The rejected child $c_2^*$ then applies to the next available daycare $d_2^*$. If daycare $d_2^*$ is also full, another child $c_3^*$ must be rejected by $d_2^*$ and apply to the subsequent daycare $d_3^*$. This sequence continues, forming a rejection chain denoted as $c_1^* \rightarrow c_2^* \rightarrow \cdots \rightarrow c_t^*$, where $t$ represents the length of the chain.

Similarly, rejection chains of families can be defined in the same manner by substituting $c_i^*$ with $f_i^*$, where $c_i^* \in C_{f_i^*}$.
\end{definition}

\begin{definition}[Rejection Cycle]
    \label{def:rejection_cycle}

A rejection chain, represented as $c_1^* \rightarrow c_2^* \rightarrow \cdots \rightarrow c_t^*$, is termed a \emph{rejection cycle} if it satisfies two additional conditions: i) at least one child in the chain is different from $c_1^*$, i.e., there exists $c' \in (c_1^*, c_2^*, \ldots, c_t^*)$ such that $c' \neq c_1^*$, and ii) the rejection chain forms a cycle, commencing and concluding with $c_1^*$, i.e., $c_1^* = c_t^*$.

In the case of a rejection cycle involving families, we mandate that i) at least two distinct families are present in the rejection chain, and ii) the rejection chain initiates and concludes with the same family. 
It is possible that the starting child $c_1^*$
 and the ending child $c_t^*$ are different, but they are from the same family.
\end{definition}

In cases where no child has siblings, rejection cycles may occur, but they are guaranteed to eventually terminate. This termination is ensured by the following reasons: i) When a daycare reaches its quota, the number of matched children remains constant, even though the set of matched children may vary. ii) Children cannot be matched to a daycare that previously rejected them, as a daycare never regrets rejecting a child with lower priority than its currently matched children when it meets its quota. Consequently, a child does not need to reapply to any daycare that has rejected them.

However, these arguments no longer hold in the presence of siblings. This is because when one child is rejected by a daycare, their sibling may be compelled to leave the matched daycare, due to their joint preferences over tuples of daycares, rather than a rejection. Consequently, vacancies arise at a daycare that was previously full, enabling a previously rejected child to reapply.
This suggests that a rejection cycle may persist indefinitely.

\subsection{Previous Algorithms}

The Sequential Couples algorithm, devised by~\citet{KPR13a} to address matching problems involving couples, follows a three-step procedure. First, it computes a stable matching without considering couples, using the DA algorithm. Next, it handles each couple according to a predefined order denoted as $\pi$. Single doctors displaced by couples are accommodated one by one, allowing them to apply to hospitals based on their preferences. However, if an application is made to a hospital where any member of a couple has previously submitted an application, the algorithm terminates unsuccessfully, even if a stable matching indeed exists.

The Sorted Deferred Acceptance (SDA) algorithm, as introduced by~\citet{ABH14a}, follows a similar trajectory to the Sequential Couples algorithm. We extend its application to the context of daycare matching with siblings. The algorithm begins by computing a stable matching without considering families with siblings, denoted as $F^S$, using the DA algorithm. Subsequently, it sequentially processes each family, denoted as $f$, based on a predefined order denoted as $\pi$. Children without siblings who are displaced by family $f$ are processed individually, enabling them to apply to daycare centers according to their preferences. If any child from family $f'\in F^S$ with siblings is affected during this process, a new order $\pi'$ is attempted, with $f$ being inserted before $f'$. The algorithm terminates if any child from family $f$ is affected or if the same permutation has been attempted twice.

In the following theorem, we demonstrate that the SDA algorithm may not produce a stable matching with respect to Definition~\ref{def:stability} when it terminates successfully. 
\begin{restatable}{theorem}{ThmSDAoriginal}\label{theo:SDA:original}
    The matching returned by the original SDA algorithm may not be stable.
\end{restatable}

\begin{proof}
We present a counterexample to prove Theorem~\ref{theo:SDA:original}.
Consider two families: $f_1$ with children $C_{f_1} = (c_1, c_2)$, $f_2$ with children $C_{f_2} = (c_3)$. 
There are four daycares: $D = \{d_0, d_1, d_2, d_3\}$, where each daycare except for $d_0$ has a single slot. The preference profile of the families and the priority profile of the daycares are as follows:
\begin{align*}
    & 
    \succ_{f_1}: (d_1, d_2), (d_2, d_3),
    \quad 
    \succ_{f_2}: d_2
    \\
    & 
    \forall d\in D, \ \succ_{d}: c_1,c_3,c_2 
\end{align*}
Then, SDA produces a matching $\mu(f_1) =\{(d_2, d_3)\}$ while leaving child $c_3$ unmatched. 
However, by Definition~\ref{def:stability}, this matching is not stable. This is because
family $f_1$ could form a blocking coalition with $(d_1, d_2)$ by allowing $c_1$ to transfer his seat at $d_2$ to sibling $c_2$.
Note that no matching for this example satisfies stability in Definition~\ref{def:stability}.
This completes the proof of Theorem~\ref{theo:SDA:original}. 
\end{proof}

\begin{algorithm*}
\caption{Extended Sorted Deferred Acceptance (ESDA)} 
\label{alg:ESDA}

\KwIn{An instance $I = (C, F, D, Q, \succ_F, \succ_D)$ and a default order $\pi = (1, 2, \dots, |F^S|)$}
\KwOut{A stable matching $\mu$ or \textit{failure}}

Let $F^O$ and $F^S$ denote the set of families with an only child and multiple children respectively \;

Initialize $\Pi \gets \{\pi\}$ (set of attempted permutations of $\pi$) \;

\For{$i \in [1, |F^S|]$}
{
    Apply Deferred Acceptance (DA) to $F^O$ and denote the resulting matching as $\mu$ \;
    Let $f$ be the $\pi(i)$-th family in $F^S$ \;

    \tcc{Proposal Step}
    Family $f$ proposes to the $p$-th tuple of daycares in its preference order, starting with $p = 0$ \;

    \tcc{Selection Step}
    Each daycare processes proposals using its choice function $Ch_d$. If any child from family $f$ is rejected, increment $p \gets p+1$ and return to the Proposal Step\;
    Otherwise, update $\mu$ by tentatively matching family $f$ to the $p$-th tuple of daycares \;

    \tcc{Check Restart Step}
    If any family $f' \in F^S$ (including $f$) has children evicted, then generate a new permutation $\pi'$ by placing $\pi(i)$ before $f'$ in the order\;
    \If{$\pi' \in \Pi$}{
        \Return \textit{failure} \;
    }
    Add $\pi'$ to $\Pi$, set $\pi \gets \pi'$, and Go back to line 3 \;

    \tcc{Stabilization Step}
    Let $RF$ denote the set of displaced families \;
    \While{$RF \neq \emptyset$}{
        Select a family $f' \in RF$ \;
        Let $f'$ repeat the Proposal, Selection, and Check Restart steps, proposing to daycares not yet examined\;
        If $f'$ is successfully matched, remove $f'$ from $RF$ \;
        Add any newly displaced families to $RF$ during this process \;
    }

    \tcc{Check Improvement Step}
    Verify whether family $\pi(i)$ can improve its assignment by allowing siblings to transfer their seats\;
    \If{an improvement is possible}{
        \Return \textit{failure} \;
    }
    Proceed to the next family: $i \gets i+1$ \;
}
\Return The matching $\mu$ \;
\end{algorithm*}

\subsection{Description of ESDA}

The ESDA algorithm extends the Sorted Deferred Acceptance (SDA) algorithm. 
We next provide an informal description of ESDA. The algorithm begins by computing a stable matching among families with an only child, denoted as $F^O$, using the Deferred Acceptance algorithm.
Subsequently, the algorithm sequentially processes each family from the set of families with multiple children, denoted as $f \in F^S$, following a predefined order $\pi$ over $F^S$.

When a family $f \in F^S$ is added to the matching process, the algorithm executes the following procedures within a single iteration.
First, in the \textbf{Proposal} step, family $f$ proposes to a tuple of daycare centers from its preference order that has not yet been considered.
Next, in the \textbf{Selection} step, each daycare evaluates these proposals using the choice function defined in Definition~\ref{def:choice_func}. If any sibling from family $f$ is rejected, the algorithm returns to the Proposal step with the next tuple in the family's preference order. Conversely, if all siblings are accepted, family $f$ is tentatively matched to the current tuple. This tentative assignment may displace some children from other families due to capacity constraints. Let $RF$ denote the set of families whose children are rejected as a result of this reallocation.
In the \textbf{Check Restart} step, if any family $f' \in F^S$ with siblings has a child rejected during this process, the algorithm attempts a new order $\pi'$ by placing $f$ before $f'$ in the sequence. If this new permutation $\pi'$ has already been attempted, the algorithm terminates and returns \textit{Unsuccess}. Otherwise, the algorithm restarts the process with $\pi'$.
Subsequently, in the \textbf{Stabilization} step, each evicted family $f' \in RF$ repeats the procedures starting from the Proposal step, proposing to the next feasible tuple in its preference order. Family $f'$ is removed from $RF$ once its assignment is determined, while any new families displaced during this process are added to $RF$. This iterative stabilization continues until $RF$ becomes empty.
Finally, in the \textbf{Check Improvement} step, the algorithm evaluates whether family $f$ can improve its current assignment by allowing siblings to transfer their seats. If such an improvement is possible, the algorithm terminates and returns \textit{Unsuccess}. Otherwise, it proceeds to process the next family in $F^S$ according to the predefined order $\pi$.

We provide a concise explanation of the differences between our ESDA algorithm and the original SDA algorithm.
First, the choice function used by daycares to select children differs significantly. In the ESDA algorithm, children can transfer their allocated seats to their siblings, a feature absent in the original SDA.
Second, the ESDA algorithm rigorously examines whether any family can form a blocking coalition with a tuple of daycares that previously rejected it, particularly when the assignment of any child without siblings is modified. In contrast, the original SDA processes each tuple of daycares only once, without performing this additional check.


\begin{example}
\label{example:ESDA}
    Consider three families $f_1$ with $C_{f_1}$ $=$ $(c_1$, $c_2)$, $f_2$ with $C_{f_2}$ $=$ $(c_3$, $c_4)$ and $f_3$ with $C_{f_3}$ $=$ $(c_5$, $c_6)$. There are five daycares denoted as $D$ $=$ $\{d_1$, $d_2$, $d_3$, $d_4$, $d_5\}$, each with one available slot. The order $\pi$ is initialized as $\{1, 2, 3\}$.
    The preference profile of the families and the priority profile of the daycares are outlined as follows:
\begin{align*}
    & \succ_{f_1}: (d_1, d_2), (d_1, d_4)  \qquad \succ_{d_1}: c_1, c_5 \qquad \succ_{d_2}: c_6, c_2 
    \\
    & \succ_{f_2}: (d_3, d_4),   (d_5, d_4)
    \qquad \succ_{d_3}: c_3, c_5 \qquad \succ_{d_4}: c_6, c_4, c_2 
    \\
    & \succ_{f_3}: (d_1, d_4),   (d_3, d_4), (d_5, d_2)
    \qquad \succ_{d_5}: c_3, c_5
\end{align*}

\textbf{Iteration 1:}
With order $\pi_1 = \{1,2,3\}$, family $f_1$ secured a match by applying to daycares $(d_1, d_2)$, followed by family $f_2$ obtaining a match with applications to $(d_3, d_4)$. However, family $f_3$ faced rejections at $(d_1, d_4)$ and $(d_3, d_4)$ before successfully securing acceptance at $(d_5, d_2)$, leading to the displacement of family $f_1$. Thus we generate a new order $\pi_2 = \{3, 1, 2\}$ by inserting $3$ before $1$.

\textbf{Iteration 2:}
With order $\pi_2 = \{3,1,2\}$, family $f_3$ secures a match at $(d_1, d_4)$. Then family $f_1$ applies to $(d_1, d_2)$ and also secures a match, resulting in the eviction of family $f_3$. This leads to the generation of a modified order $\pi_3 = \{1, 3, 2\}$ with $1$ preceding $3$.

\textbf{Iteration 3:}
With order $\pi_3 = \{1, 3, 2\}$, family $f_1$ secures a match at $(d_1, d_2)$. Subsequent applications by $f_3$ result in a match at $(d_3, d_4)$, but $f_2$ remains unmatched due to rejections at $(d_3, d_4)$ and 
$(d_5, d_4)$. The algorithm terminates, returning a stable matching $\mu$ with $f_1$ matched to $(d_1, d_2)$ and $f_3$ matched to $(d_3, d_4)$, while $f_2$ remains unmatched.
\end{example}

\subsection{Two Types of Unsuccessful Termination in ESDA}
\label{sec:unsuccessful_types}

The ESDA algorithm terminates unsuccessfully in two scenarios suggesting that a stable matching may not exist, even if one indeed exists.

A \emph{Type-1 Unsuccessful Termination} happens when, during the insertion of a family $f \in F^S$, a child $c \in C_f$ initiates a rejection chain that ends with another child $c' \in C_f$ from the same family, where all other children in the chain do not have siblings. This unsuccessful termination is further divided into two cases based on whether $c = c'$ holds: Type-1-a Unsuccessful Termination when $c=c'$ and Type-1-b Unsuccessful Termination when $c\neq c' \in f_c$.

\begin{example}[Type-1-a Unsuccessful Termination]
\label{example:type-1a}
Consider three families $f_1$ with children $C_{f_1} = (c_1, c_2)$, $f_2$ with children $C_{f_2} = \{c_3\}$ and $f_3$ with children $C_{f_3} = \{c_4\}$. There are three daycares denoted as $D=\{d_1, d_2, d_3\}$, each with one available slot. The preferences of the families and the priorities of the daycares are outlined as follows:
\begin{align*}
    & \succ_{f_1}: (d_1, d_3) \quad \succ_{f_2}: d_1, d_2 \quad \succ_{f_3}: d_2, d_1
    \\
    & \succ_{d_1}: c_4, c_1, c_3 \quad \succ_{d_2}: c_3, c_4 \quad \succ_{d_3}: c_2
\end{align*}
The initial matching $\mu^O$ is obtained through the Deferred Acceptance (DA) algorithm, where $\mu^O(c_3) = d_1$ and $\mu^O(c_4) = d_2$. Upon inserting family $f_1$, child $c_1$ is matched to daycare $d_1$, and child $c_2$ is matched to daycare $d_3$, resulting in the rejection of child $c_3$ from daycare $d_1$. Subsequently, when child $c_3$ applies to daycare $d_2$, it leads to the rejection of child $c_4$. Finally, when child $c_4$ applies to daycare $d_1$, it results in the rejection of child $c_1$.

Thus, a rejection chain is formed: $c_1 \rightarrow c_3 \rightarrow c_4 \rightarrow c_1$, and the ESDA algorithm terminates unsuccessfully. However, it's important to note that a stable matching $\mu'$ does exist, where $\mu'(c_3) = d_2$ and $\mu'(c_4) = d_1$. Despite its existence, both the SDA and the ESDA algorithms fail to discover it.
\end{example}

\begin{example}[Type-1-b Unsuccessful Termination]
\label{example:type-1b}
Consider two families $f_1$ with children $C_{f_1} = (c_1, c_2)$ and $f_2$ with children $C_{f_2} = \{c_3\}$. There are two daycares $D=\{d_1, d_2\}$, each having one available slot. The preferences of the families and the priorities of the daycares are outlined as follows:
\begin{align*}
    & \succ_{f_1}: (d_1, d_2) \quad \succ_{f_2}: d_1, d_2
    \\
    & \succ_{d_1}: c_1, c_3 \quad \succ_{d_2}: c_3, c_2
\end{align*}
The initial matching $\mu^O$ is obtained through the Deferred Acceptance (DA) algorithm, with $\mu^O(c_3) = d_1$. Upon the introduction of family $f_1$, child $c_1$ secures a place at daycare $d_1$, and child $c_2$ is matched with daycare $d_2$, consequently leading to the rejection of child $c_3$ from daycare $d_1$. As child $c_3$ applies to daycare $d_2$, it results in the rejection of child $c_2$ from daycare $d_2$ in turn.

This sequence forms a rejection chain: $c_1 \rightarrow c_3 \rightarrow c_2$, prompting the ESDA algorithm to terminate unsuccessfully. Notably, no stable matching is found to exist for Example~\ref{example:type-1b}.
\end{example}

A \emph{Type-2 Unsuccessful Termination} occurs when two families, $f_1$ and $f_2 \in F^S$, satisfy the following conditions:
i) $f_1$ precedes $f_2$ in the current order $\pi$,
ii) There exists a rejection chain starting from $f_2$ and ending with $f_1$, where all other families in the chain have only one child, and iii) A new order $\pi'$ is generated by placing $f_2$ before $f_1$, and this order has been attempted and stored in the set $\Pi$, which keeps track of permutations explored during the ESDA process.

\begin{example}[Type-2 Unsuccessful Termination]
\label{example:type-2}
Consider two families $f_1$ with children $C_{f_1} = (c_1, c_2)$, and $f_2$ with children $CC_{f_2} = (c_3,c_4)$. There are three daycares, denoted as $D=\{d_1, d_2, d_3\}$, each with one slot. Suppose the initial order is $\pi = \{1, 2\}$. The preferences of the families and the priorities of the daycares are outlined as follows:
\begin{align*}
    & \succ_{f_1}: (d_1, d_2), (d_1, d_3) \quad \succ_{f_2}: (d_2, d_3)
    \\
    & \succ_{d_1}: c_1 \quad \succ_{d_2}: c_3, c_2 \quad \succ_{d_3}: c_2, c_4
\end{align*}

When family $f_1$ is inserted, it secures a match with $(d_1, d_2)$. Subsequently, when family $f_2$ is added, child $c_2$ from family $f_1$ is rejected, prompting a change in the order to $\pi' = \{2, 1\}$ and a restart of the algorithm.
Now, if we add family $f_2$ first in the revised order $\pi'$, it obtains a match with $(d_2, d_3)$. However, when family $f_1$ is added and applies to $(d_1, d_2)$, child $c_2$ has a lower priority than child $c_3$, resulting in the rejection of family $f_1$. Consequently, family $f_1$ applies to $(d_1, d_3)$, causing family $f_2$ to be evicted in turn.
This leads us to modify the order $\pi'$ to $\pi^* = \{1, 2\}$, which has been attempted previously. Thus, the ESDA algorithm terminates due to a Type-2 Unsuccessful Termination.
\end{example}

These two types of unsuccessful terminations are crucial when analyzing the probability of the existence of stable matchings in a large random market.

\subsection{Successful Termination}
\label{subsec:ESDA:success}

We next demonstrate that ESDA always generates a stable matching if it terminates successfully. 

\begin{restatable}{theorem}{ThmESDAstability}\label{theo:ESDA:stability}
    Given an instance of $I$, if ESDA returns a matching, then the yielded matching is stable. In addition, ESDA always terminates in a finite time.
\end{restatable}


Our proof that ESDA always generates a stable matching if it terminates successfully, 
relies on the following two lemmas. First, we establish that the number of matched children at each daycare does not decrease as long as no family in $F^S$ is rejected and no child passes their seat to other siblings during the execution of ESDA. Then, we prove that for a given order $\pi$ over $F^S$, if the rank of the matched child at any daycare increases, then ESDA cannot produce a matching with respect to $\pi$.

\begin{lemma}
\label{lemma:ESDA:decrease}
For a given order $\pi$ over families $F^S$, let $\mu^i(\pi)$ denote the matching obtained during the ESDA procedure before processing the $i$-th family denoted as $F^S_{\pi(i)} \in F^S$. The number of matched children at any daycare $d$ does not decrease under matching $\mu^{i+1}(\pi)$ if the following three conditions hold: 
i) The algorithm does not encounter any type of Unsuccessful Termination.
ii) The order $\pi$ remains unchanged.
iii) No child from family $F^S_{\pi(i+1)}$ transfers their seat to other siblings during the ESDA process.
\end{lemma}

\begin{proof}
If the first two conditions hold, then no child from any family $f \in F^S$ is rejected when inserting family $F^S_{\pi(i+1)}$. Consequently, only children without siblings are involved in rejection chains, and each time one child is replaced by another one with a higher daycare priority when the capacity is reached.

Let $f = F^S_{\pi(i+1)}$. If the third condition holds, when family $f$ applies to any tuple of daycares $D(\succ_f, j)$, the input to the choice function $\Ch_d(\cdot)$ can be simplified as $\Ch_d\bigl(\mu(d) \cup C(\succ_f, j, d)\bigr)$, as no child $c \in C_f$ passes their seat to other siblings. After the stabilization step, if $f$ reapplies to any tuple $\succ_{f, k}$ that is better than $\mu(f)$, then $f$ is still rejected as each matched child at $d\in D(\succ_f, j)$ has a weakly higher priority. Thus, $f$ cannot create new vacancies by moving to a better tuple of daycares. Consequently, the number of matched children at each daycare does not decrease.
\end{proof}

For a given matching $\mu$ and a daycare $d$, let $Rank(\mu, d)$ represent the rank of the matched child with the lowest priority at daycare $d$, where $1$ denotes the highest priority. Imagine that all vacant slots at each daycare are initially occupied by dummy children assigned the rank $|C|+1$. As the ESDA algorithm progresses, these dummy children are gradually rejected and replaced by children with higher priorities, resulting in a decrease in $Rank(\cdot)$.

We will now demonstrate the following lemma.

\begin{lemma}
\label{lemma:ESDA:rank}
Given an order $\pi$ over families $F^S$, if, during the ESDA process, $Rank(\mu, d)$ increases for any daycare $d$,  then ESDA fails to generate a matching under the current order $\pi$ over families $F^S$.
\end{lemma}
\begin{proof}
\label{proof:lemma:ESDA:rank}
We proceed to prove Lemma~\ref{lemma:ESDA:rank} by analyzing how $Rank(\mu, d)$ evolves for each daycare $d$ during the execution of the ESDA algorithm under the specified order $\pi$.

The ESDA algorithm begins by applying the DA algorithm to the families in $F^O$. In each step of the DA algorithm, any rejected child is replaced by another child with higher priority. As a result, the value of $Rank(\mu, d)$ for each daycare $d$ either decreases or remains unchanged.

Next, the algorithm proceeds through the families in $F^S$ according to the specified order $\pi$. Consider the insertion of family $f = F^S_{\pi(i)}$ into the market, starting with $i \gets 1$. The following argument applies to any $i$, provided that no child from family $F^S_{\pi(i)}$ transfers their seat to a sibling.

In the Proposal and Selection Step, family $f$ begins by applying to the tuple of daycares $D(\succ_f, p)$, with $p$ initialized to 1. If family $f$ is not accepted by all $d \in D(\succ_f, p)$, the set of matched children at each daycare $d$ remains unchanged, meaning $Rank(\mu, d)$ stays the same. The algorithm then proceeds to the next tuple with $p \gets p+1$. If $D(\succ_f, j)$ can accommodate family $f$, then children with lower priority are replaced by the members of $C_f$. This substitution causes a decrease in $Rank(\mu, d)$ for each daycare $d \in D(\succ_f, j)$.

In the Check Restart Step, two possible situations arise in this scenario.

\textbf{Case (i):} If a child from another family $f' \in F^S$ is rejected, this can either trigger a restart with a new permutation or lead to an Unsuccessful Termination. Notably, if $f = f'$ (i.e., a child from family $f$ is evicted), it still results in an Unsuccessful Termination, as the updated permutation remains unchanged. In either case, the situation can be viewed as filling all seats at each daycare with dummy children assigned the rank $|C|+1$, causing an increase in $Rank(\cdot)$. This outcome indicates that the current order $\pi$ is incapable of producing a valid matching.

\textbf{Case (ii):} If only children in $C^O$ are affected during the insertion of $f$, family $f$ is matched to $D(\succ_f, j)$, and any evicted child is assigned to the dummy daycare. In this case, $Rank(\cdot)$ decreases at each daycare $d \in D(\succ_f, j)$. Furthermore, if a child is rejected, they are replaced by another child with a higher priority, which also results in a decrease in $Rank(\cdot)$ at the corresponding daycare.

In the Stabilization Step, we process rejected families by repeating the above steps, and the analysis remains the same.

In the Check Improvement Step, if family $f$ reapplies to $D(\succ_f, k)$ after being rejected and is successfully matched, it must be because a child $c$ transfers their seat to their sibling $c'$, resulting in an increase in $Rank(\mu, d)$ at some daycare $d$. However, this implies that $c'$ was initially rejected by $d$, where another child $c^*$—who had the lowest priority among those matched to $d$—has a higher priority than $c'$. Consequently, the updated matching cannot be stable, as the evicted child $c^*$ has a legitimate claim to $d$ over $c'$. As a result, the ESDA algorithm terminates unsuccessfully without re-entering the Check Restart Step.

After meticulously analyzing all possible scenarios during the ESDA procedure, it becomes clear that $\pi$ cannot result in a matching if $Rank(\mu, d)$ increases for any daycare $d$. This concludes the proof of Lemma~\ref{lemma:ESDA:rank}.
\end{proof}

We now present a formal proof of Theorem~\ref{theo:ESDA:stability}.


\begin{proof}[Proof of Theorem~\ref{theo:ESDA:stability}]
Suppose the ESDA in Algorithm~\ref{alg:ESDA} returns a matching $\mu$ successfully. Let $\tilde{\pi}$ denote the final order over families $F^S$ when ESDA terminates.

Let $w = |F^S|$ denote the number of families in $F^S$, and consider the last family $f^w = F^S_{\tilde{\pi}(w)}$ in the order $\tilde{\pi}$.

\textbf{Case i):} If family $f^w$ is matched to $\mu(f) = D(\succ_f, j)$ without causing any child to be rejected (i.e., the stabilization step is not invoked), then for any $k \leq j$, family $f$ cannot be matched to a better tuple of daycares $D(\succ_f, k)$, as the set of matched children remains unchanged at any $d \in D(f, k)$.

\textbf{Case ii):} Suppose some families $RF$ are rejected when inserting family $f^w$. We know that $RF \setminus F^S = \emptyset$, otherwise ESDA would terminate unsuccessfully or restart with a new permutation. Thus, $RF \subseteq F^O$. After stabilizing all families in $RF$, we check whether family $f$ can be improved by reapplying to a better tuple of daycares, allowing for children in $C_f$ to pass their seats to other siblings. If this happens, the rank of matched children $Rank(\cdot)$ at some daycare decreases, contradicting Lemma~\ref{lemma:ESDA:rank}, which implies that $\tilde{\pi}$ can produce a matching. Therefore, we conclude that $f$ cannot be matched to a better tuple.

For both cases, we conclude that family $f^w$ cannot participate in a blocking coalition with respect to the matching $\mu$.

Moving on to the second last family $f^{w-1}$, we apply similar reasoning. When inserting family $f^{w-1}$ into the market, if it can be matched to a better tuple after the stabilization step, it would contradict Lemma~\ref{lemma:ESDA:rank}. After family $f^w$ is introduced into the market, two key observations hold: 
\begin{itemize}
    \item [(i)] The number of matched children does not decrease at any daycare, as per Lemma~\ref{lemma:ESDA:decrease}, and
    \item [(ii)] For each daycare $d$, $Rank(\mu, d)$ does not increase, meaning no daycare accepts a child with a lower priority, per Lemma~\ref{lemma:ESDA:rank}.
\end{itemize}
Consequently, family $f^{w-1}$ still cannot be matched to a better tuple of daycares after the insertion of the last family $f^w$. 
Continuing this logic through induction, we conclude that no family $f^i \in F^S_{\pi(i)}$ can be matched to a better tuple of daycares under the order $\tilde{\pi}$. In other words, none of the families in $F^S$ can participate in a blocking coalition. 
For the same reasons, it follows that any family $f \in F^O$ cannot be matched to a better daycare either.

For each permutation of $\pi$, the algorithm only backtracks to check whether the currently processed family $f \in F^S$ can be better off. If this happens, the algorithm terminates unsuccessfully. Otherwise, it moves on to the next family. 
Since the choices in each family's preference ordering are finite, the check terminates in finite time. Furthermore, the total number of permutations of $\pi$ is finite, ensuring the algorithm's termination.

This concludes the proof of Theorem~\ref{theo:ESDA:stability}.
 \end{proof}

\begin{table*}[H]
\centering
\caption{Important notations for random daycare market.}
\label{tab:notation}
\begin{tabular}{c|c}
\hline
$\alpha \in [0,1]$ & Percentage of children with siblings \\
$K \ge 1$ & Maximum number of siblings within each family \\
$L \ge 1$ & Maximum length of individual preference orderings \\
$\sigma \ge 1$ & Parameter in the uniformly bounded condition \\
$\varepsilon \ge 0$ & Parameter used in generating $\succ_0$ \\
$\phi \in (0,1]$ & Dispersion parameter used in a Mallows model \\
\hline
\end{tabular}
\end{table*}

\section{Random Daycare Market}
To analyze the likelihood of a stable matching in practice, we proceed to introduce a random market where preferences and priorities are generated from probability distributions.
Formally, we represent a random daycare market as $\tilde{I} = (C, F, D, Q, \alpha, K, L, \mathcal{P}, \rho, \sigma, \mathcal{D}_{\succ_0, \phi}, \varepsilon)$. 

Let $|C| = n$ and $|D| = m$ denote the number of children and daycares, respectively. Throughout this paper, we assume that
$m=\Omega(n)$. 
To facilitate analysis, we partition the set \(F\) into two distinct groups labeled \(F^S\) and \(F^O\), signifying the sets of families with and without multiple children, respectively. Correspondingly, \(C^S\) and \(C^O\) represent the sets of children with and without siblings, respectively.
The parameter $\alpha$ signifies the percentage of children with siblings.  Then we have $|C^O| = (1-\alpha)n$ and $|C^S| = \alpha n$.
For each family $f$, the size of $C_f$ is constrained by a constant $K$, expressed as $\max_{f \in F} |C_f| \leq K$.

\subsection{Preferences of Families}

We adopt the approach described in~\citep{KPR13a} to generate family preferences using a two-step process. In the first step, we independently generate preference orderings for each child based on a probability distribution $\mathcal{P}$ over the set of daycares $D$.
Let $\mathcal{P} = (p_d)_{d \in D}$ denote a probability distribution, where $p_d$ represents the probability of selecting daycare $d$. For each child $c$, we initialize an empty list, then independently select a daycare $d$ from $\mathcal{P}$ and add it to the list if it is not already included. This process is repeated until the list reaches a maximum length $L$, which is typically a small constant in practice~\citep{STM+23a}.

We adhere to the assumption that the distribution $\mathcal{P}$ satisfies a \emph{uniformly bounded} condition, as assumed in the random market by~\citep{KPR13a} and~\citep{ABH14a}.

\begin{definition}[Uniformly Bounded]
\label{def:uniformly_bounded}
For all $d, d' \in D$, the ratio of probabilities $p_d / p_{d'}$ falls within the interval $[1/\sigma, \sigma]$ with a constant $\sigma \geq 1$. 
\end{definition}

\begin{restatable}{lemma}{LemmaUniform}\label{lemma:uniform_bound}
Under the uniformly bounded condition, the probability $p_d$ of selecting any daycare $d$ is limited by $\sigma/m$ where $m$ denotes the total number of daycares.
\end{restatable}

\begin{proof}
For  all $d, d' \in D$, we have $1/\sigma \leq p_d / p_{d'} \leq \sigma$. Therefore, $ p_{d'} / \sigma \leq p_d \leq \sigma \cdot p_{d'}$. If we sum this inequality over each $d' \in D$, we obtain $m \cdot p_d \leq \sum_{d' \in D} \sigma \cdot p_{d'} = \sigma$. Thus, $p_d \leq \sigma/m$.
\end{proof}

In the second step, we generate all possible combinations of the individual preferences of the children within each family. From these combinations, we uniformly at random select a subset with a specified length limit, without imposing additional restrictions.

\subsection{Priorities of Daycares}
A notable departure from previous work~\citep{KPR13a,ABH14a} is our adoption of the Mallows model~\citep{Mall57a} to generate daycare priority orderings over children. 
The Mallows model, denoted as $\mathcal{D}_{\succ_0, \phi}$, begins with a reference ordering $\succ_0$. New orderings are then probabilistically generated based on this reference, with the degree of deviation controlled by a dispersion parameter $\phi$. 

Through active collaborations with multiple Japanese municipalities, we have observed that daycare centers often share similar priority structures for children. In practice, municipalities typically use a complex scoring system to assign a unique priority score to each child, establishing a strict priority order. This order is then applied and slightly adjusted by each daycare based on their individual policies (e.g., prioritizing siblings who are already enrolled). The Mallows model is particularly well-suited for replicating these priority orderings, as it allows for controlled variations around a reference ranking. Additionally, this model is widely recognized for its flexibility and has been extensively used for preference generation across various domains~\citep{LuCr11a,BrHo22a}.

Let $S$ denote the set of all orderings over $C$. 
\begin{definition}[Kendall-tau Distance]
For a pair of orderings $\succ$ and $\succ'$ in $S$, the \emph{Kendall-tau distance}, denoted by $\mathrm{inv}(\succ, \succ')$, is a metric that counts the number of pairwise inversions between these two orderings. Formally,
$
    \mathrm{inv}(\succ,\succ') = |\{ c,c' \in C \mid c \succ' c' \text{ and } c' \succ c \}|.
$
\end{definition}
\begin{definition}[Mallows Model]
Let $\phi \in (0,1]$ be a dispersion parameter and $Z= \sum_{\succ \in S} \phi^{\mathrm{inv}(\succ,\succ_0)}$. The \emph{Mallows distribution} is a probability distribution over $S$. The probability that an ordering $\succ$ in $S$ is drawn from the Mallows distribution is given by
\[
    \mathrm{Pr}[\succ] = \frac{1}{Z}\, \phi^{\mathrm{inv}(\succ,\succ_0)}.
\]
\end{definition}
The dispersion parameter $\phi$ characterizes the correlation between the sampled ordering and the reference ordering $\succ_0$. Specifically, when $\phi$ is close to $0$, the ordering drawn from $\mathcal{D}_{\succ_0,\phi}$ is almost the same as the reference ordering $\succ_0$. On the other hand, when $\phi=1$, $\mathcal{D}_{\succ_0,\phi}$ corresponds to the uniform distribution over all permutations of $C$.

Siblings within the same family typically share identical priority scores, with ties resolved arbitrarily~\citep{STM+23a, SYT+24a}. Motivated by this observation, we construct a reference ordering $\succ_0$ through the following steps. Starting with an empty list, we first include all singleton children $C^O$, who do not have siblings. For each family $f \in F^S$ (families with siblings), we decide probabilistically whether to add its children individually or as a grouped entity: with a probability of $1/n^{1+\varepsilon}$, all children $C_f$ of the family are added as separate entries, and with a probability of $1 - 1/n^{1+\varepsilon}$, the family is added as a single entity to keep its children grouped together, where $n$ is the total number of children and $\varepsilon > 0$ is a constant. Once all children and families are added, the list is shuffled to introduce randomness. Finally, the reference ordering $\succ_0$ is drawn from a uniform distribution over all permutations of the shuffled list.

\begin{example}[Generate Reference Ordering $\succ_0$]
    \label{example:generate_reference_ordering}
Suppose there are five children: $C = \{c_1, c_2, c_3, c_4, c_5\}$, where $c_1$, $c_2$, and $c_3$ belong to family $f_1$, and $c_4$ and $c_5$ are singletons. We start with an empty list and add the singleton children $c_4$ and $c_5$, resulting in $[c_4, c_5]$. For family $f_1$, we probabilistically choose between adding its children individually, e.g., $[c_4, c_5, c_1, c_2, c_3]$, or as a grouped entity, e.g., $[c_4, c_5, f_1]$. Suppose we choose the latter. Then, the list is shuffled randomly, for example, $[c_4, f_1, c_5]$. A uniform distribution over all permutations of this shuffled list is used to generate $\succ_0$, such as $\succ_0 = c_4 \succ c_5 \succ c_1 \succ c_2 \succ c_3$, where all children from $f_1$ are grouped together in the generated ordering.  

In other words, with a probability of $1/n^{1+\varepsilon}$, we treat siblings from the same family separately, and with a probability of $1-1/n^{1+\varepsilon}$, we treat them as a whole, or more precisely, as a continuous block in $\succ_0$.
\end{example}

\begin{definition}[Diameter]
Given a reference ordering $\succ_0$ over children $C$, we define the \emph{diameter} of family $f$, denoted by $\mathrm{diam}_f$, as the greatest difference of positions in $\succ_0$ among $C_f$ plus 1. Formally,
\[
\mathrm{diam}_f = \mathrm{position}\left(\max_{c'\in C_f}\right) - 
\mathrm{position}\left(\min_{c'\in C_f}\right) + 1,
\]
where $\max_{c\in C_f} c$ (resp. $\min_{c\in C_f} c$) refers to the child in $C_f$ with the highest (resp. lowest) priority in $\succ_0$. 
\end{definition}

The methodology employed to generate the reference ordering $\succ_0$ above adheres to the following condition.
For each family $f$ with siblings, we have
$
    \mathrm{Pr}\bigl[\mathrm{diam}_f \geq |C_f|\bigr] \leq \frac{1}{n^{1+\varepsilon}}
$
from the construction.

\subsection{Main Theorem}
We focus on a random market $\tilde{I}$ where all parameters are set as described above. While Example~\ref{example:non_existence_stable} demonstrates that a stable matching may not exist even when all daycares have the same priority ordering over children, our main result, encapsulated in the following theorem, shows that for a large random market, the existence of a stable matching is highly likely.

\begin{restatable}{theorem}{ThmMain}\label{theorem:main}
Given a random market $\tilde{I}$ with $\phi = O(\log n/n)$, the probability of the existence of a stable matching converges to $1$ as $n$ approaches infinity.
\end{restatable}

\begin{table}[H]
\centering
\caption{Important notations for random daycare market.}
\label{tab:notation}
\begin{tabular}{cc}
\hline
$\alpha$ & Percentage of siblings \\
$\beta$ & Maximum number of siblings within each family \\
$L$ & Maximum length of individual preference orderings \\
$\mathcal{P}$ & Probability distribution over daycares $D$ \\
$\rho$ & Function aggregating individual preferences \\
$\sigma$ & Parameter in the uniformly bounded condition \\
$\mathcal{D}_{\succ_0, \phi}$ & Mallows Model with a reference ordering $\succ_0$ and a dispersion parameter $\phi$ \\
$\varepsilon$ & Parameter used in generating $\succ_0$ \\
\hline
\end{tabular}
\end{table}

We will prove Theorem~\ref{theorem:main} by showing that the Extended Sorted Deferred Acceptance (ESDA) algorithm produces a stable matching with a probability that converges to 1 in the random market. Our primary approach to proving Theorem~\ref{theorem:main} involves setting an upper bound on the likelihood of encountering the two types of unsuccessful termination in the ESDA algorithm.

The following two lemmas establish that as $n$ approaches infinity, Type-1-a and Type-1-b unsuccessful terminations are highly unlikely to occur when the dispersion parameter $\phi$ is on the order of $O(\log n/n)$. We defer the proofs for these results to Appendices~\ref{appendix:proof_of_type_1_a} and~\ref{appendix:proof_of_type_1_b}, respectively.
\begin{restatable}{lemma}{LemmaTypeOneA}\label{lemma:type-1-a}
    Given a random market $\tilde{I}$ with $\phi = O(\log n/n)$,  the probability of Type-$1$-a unsuccessful termination in the ESDA algorithm is bounded by 
    $O\bigl((\log n)^2/n\bigr)$.
\end{restatable}

\begin{restatable}{lemma}{LemmaTypeOneB}\label{lemma:type-1-b}
    Given a random market $\tilde{I}$ with $\phi = O(\log n/n)$, the probability of Type-$1$-b unsuccessful termination in the ESDA algorithm is bounded by 
    $O\bigl((\log n)^2/n\bigr) + O(n^{-\varepsilon})$.
\end{restatable}

As illustrated in Examples~\ref{example:non_existence_stable}, Type-2 unsuccessful termination can occur even when the priorities of daycares over children are identical.

 We introduce concepts of \emph{domination} and \emph{nesting} to analyze the case of Type-2 unsuccessful termination. 
\begin{definition}[Domination]
\label{def:dominiation}
Given a priority ordering $\succ$, we say that family $f$ \emph{dominates} $f'$ w.r.t. $\succ$ if 
$
\max_{c\in C_f} c \succ \min_{c'\in C(f')} c'
$
where $\max_{c\in C_f} c$ (resp. $\min_{c\in C_f} c$) represents the child in $C_f$ with the highest (resp. lowest) priority under the priority ordering $\succ$. 
\end{definition}
In simple terms, if $f$ dominates $f'$, then there is a possibility that $f'$ will be rejected by daycares with a certain order $\succ$ due to an application of $f$.

\begin{definition}[Top Domination]
\label{def:top_dom}
Given a priority ordering $\succ$, we say that family $f$ \emph{top-dominates} $f'$ w.r.t. $\succ$ if 
\[
   \max_{c\in C_f} c \succ \max_{c'\in C(f')} c'.
\] 
\end{definition}

Intuitively, a Type-2 unsuccessful termination can arise from a cycle in which two families with siblings reject each other. We introduce the concept of \emph{nesting} as follows.
\begin{definition}[Nesting]
\label{def:nest}
    Given a priority ordering $\succ$, two families $f$ and $f'$ are said to be \emph{nesting} if they mutually dominate each other under $\succ$.
\end{definition}

\begin{example}
\label{example:nest}
Consider three families $F=\{f_1,f_2,f_3\}$, each with two children: $C_{f_1}=(c_1,c_2)$, $C_{f_2}=(c_3,c_4)$, and $C_{f_3}=(c_5,c_6)$. Suppose there is a priority ordering $\succ$: $c_1$, $c_3$, $c_5$, $c_2$, $c_4$, $c_6$. In this case, all pairs in $F$ nest with each other with respect to $\succ$. 
\end{example}

We next show that if any two families do not nest with each other with respect to $\succ_0$, then Type-2 unsuccessful termination is unlikely to occur as $n$ tends to infinity in Lemma~\ref{lem:type-2-lem}. We defer the proof to Appendix~\ref{appendix:proof_lemma_type_2_lem}.

\begin{restatable}{lemma}{LemmaTypeTwoLemma}\label{lem:type-2-lem}
    Given a random market $\tilde{I}$ with $\phi = O(\log n/n)$, and for any two families $f, f'\in F^S$ that are not nesting with each other with respect to $\succ_0$, then Type-$2$ unsuccessful termination occurs with a probability of at most $O(\log n / n)$.
\end{restatable}

Following an analysis of the probability that any two pairs of families from $F^S$ nest with each other with respect to the reference ordering $\succ_0$, we establish the probability of Type-$2$ unsuccessful termination in Lemma~\ref{lemma:type-2}.

\begin{restatable}{lemma}{LemmaTypeTwo}\label{lemma:type-2}
     Given a random market $\tilde{I}$ with $\phi = O(\log n/n)$, the probability of Type-$2$ unsuccessful termination occurring is bounded by $O(\log n / n) + O\bigl(n^{-2\varepsilon}\bigr)$.
\end{restatable}

Lemma~\ref{lemma:type-1-a}, Lemma~\ref{lemma:type-1-b}, and Lemma~\ref{lemma:type-2} imply the existence of a stable matching with high probability for the large random market, thus concluding the proof of Theorem~\ref{theorem:main}. 

\section{Formal Proof of Theorem~\ref{theorem:main}}

In this section, we present a detailed proof for Theorem~\ref{theorem:main}. 

We leverage the following lemma in our proof. It asserts that if an ordering $\succ$ is generated from a given Mallows distribution $\mathcal{D}_{\succ_0,\phi}$, the probability of child $c'$ being ranked higher than child $c$ in $\succ$ is no greater than $4 \phi^{\mathrm{dist}(c,c')}$, given that $c \succ_0 c'$, where $\mathrm{dist}(c,c')$ represents the distance between $c$ and $c'$ in $\succ_0$. 
\begin{lemma}[\cite{Levy17a}]\label{lem:levy}
If $\succ$ is a random ordering drawn from the Mallows distribution $\mathcal{D}_{\succ_0,\phi}$, then for all $c,c' \in C$,
\[
    \mathrm{Pr}\bigl[c' \succ c \mid c \succ_0 c'\bigr] \leq 4 \phi^{\mathrm{dist}(c,c')},
\]
where $\mathrm{dist}(c,c') = |\{c'' \in C \mid c \succ_0 c'' \succ_0 c'\}|+1$.
\end{lemma}


\subsection{Proof of Lemma~\ref{lemma:type-1-a}}\label{appendix:proof_of_type_1_a}

\LemmaTypeOneA*

\begin{proof}
We first consider a Type-1-a unsuccessful termination, where a rejection chain $c_1 \rightarrow c_2^* \rightarrow \cdots \rightarrow c_{\ell}^* \rightarrow c_1$ exists. Here, child $c_1$ belongs to a family $f \in F^S$ with multiple children, while the other children $c_2^*,c_2^*, \ldots, c_{\ell}^* \in C^O$ have no siblings.

Let $\mathcal{E}_{\ell}^{\mathrm{a}}$ represent the event of such a rejection chain $c_1 \rightarrow c_2^* \rightarrow \cdots \rightarrow c_{\ell}^* \rightarrow c_1$, with length $\ell\geq 3$.
We next show that, for any $\succ_0$, we have 
\begin{align}\label{eq:bound_for_E}
    \mathrm{Pr}[\mathcal{E}_{\ell}^{\mathrm{a}} \mid \succ_0] \leq \frac{16\sigma \phi^2}{m}.
\end{align}

Suppose that in this rejection chain, child $c_1$ applies to daycare $d_1$, while children $c_i^*$ apply to $d_i^*$ for $i \in \{2,3,...,\ell-1\}$. The last child in the cycle, $c_{\ell}^*$, applies to daycare $d_1$. 
It is important to note that $d_i^* \neq d_{i+1}^*$ holds for $i \in \{1,2,\ldots,\ell-2\}$, even though there could be repetitions among the children $c_2^*,c_3^*,\ldots,c_{\ell}^*$ and the daycares $d_2^*,d_3^*,\ldots,d_{\ell-1}^*$.

Let $\succ_1$ represent the priority ordering of daycare $d_1$. For $i \in \{2,3,\ldots,\ell-1\}$, let $\succ_i$ denote the priority ordering of daycare $d_i^*$. Recall that for each $i=1,2,\ldots,\ell-1$, the priority ordering $\succ_i$ is drawn from the Mallows distribution $\mathcal{D}_{\succ_0,\phi}$. We consider two cases.

Case (i): Suppose the reference ordering $\succ_0$ satisfies the following condition
\begin{equation}
\label{equation:type1a}
c_{\ell}^*\succ_0c_{\ell-1}^*\succ_0\cdots \succ_0 c_2^*\succ_0 c_1.
\end{equation} 
By Lemma~\ref{lem:levy}, we have 
\begin{equation*}
    \mathrm{Pr}[c_{\ell}^* \succ_1 c_1 \succ_1 c_2^* \mid \succ_0] \leq \mathrm{Pr}[c_1 \succ_1 c_2^* \mid c_2^* \succ_0 c_1]\leq 4\phi.
\end{equation*}
For all $i=2,3,\ldots,\ell-1$, we also have
\begin{equation*}
    \mathrm{Pr}[c_{i}^* \succ_{i} c_{i+1}^*\mid \succ_0] \leq 4\phi. 
\end{equation*}
From $d_1^* \neq d_{2}^*$, we know $\succ_1$ and $\succ_{2}$ are independent. Then we have
\begin{align*}
    \mathrm{Pr}\bigl[\mathcal{E}_{\ell}^{\mathrm{a}} \mid \succ_0 \bigr]  
    &\leq \mathrm{Pr}\bigl[c_{1} \succ_{1} c_{2}^* \mid \succ_0 \bigr]\cdot \mathrm{Pr}\bigl[c_{2}^* \succ_{2} c_{3}^* \mid \succ_0 \bigr]
    \cdot 
    \mathrm{Pr}\bigl[\text{$c_{\ell-1}^*$ applies to $d_1$}\bigr] \\
    &\leq 16\phi^2 p_{d_1}.
\end{align*}
Lemma~\ref{lemma:uniform_bound} states that $p_{d_1}\leq \sigma/m$. Then we have 
\begin{equation}\label{eq:case_1_lemma_2}
    \mathrm{Pr}\bigl[\mathcal{E}_{\ell}^{\mathrm{a}} \mid \succ_0 \bigr] 
    \leq 16\phi^2 p_{d_1} \leq \frac{16\sigma \phi^2}{m}. 
\end{equation}

Case (ii): If $\succ_0$ does not satisfy the condition in Formula~\eqref{equation:type1a},
 then $\mathrm{Pr}[c_{\ell}^* \succ_1 c_1 \succ_1 c_2^* \mid \succ_0]\leq 4\phi^2$ holds or there exists $i\in \{2,3,\ldots,\ell-1\}$ such that $\mathrm{Pr}[c_{i}^* \succ_{i} c_{i+1}^* \mid \succ_0] \leq 4\phi^2$. 
From this, we obtain 
\begin{align}\label{eq:case_2_lemma_2}
    \mathrm{Pr}\bigl[\mathcal{E}_{\ell}^{\mathrm{a}} \mid \succ_0 \bigr] \nonumber
    &\leq 4\phi^2\cdot \mathrm{Pr}\bigl[\text{$c_{\ell-1}^*$ applies to $d_1$}\bigr] \\&\leq 4\phi^2 p_{d_1} \nonumber \\
    &\leq \frac{4\sigma \phi^2}{m}. 
\end{align}
From Inequalities~\eqref{eq:case_1_lemma_2} and~\eqref{eq:case_2_lemma_2} above, for both cases (i) and (ii), we have $\mathrm{Pr}[\mathcal{E}_{\ell}^{\mathrm{a}} \mid \succ_0] \leq \frac{16\sigma \phi^2}{m}$.
This completes the proof of Inequality~\eqref{eq:bound_for_E}.

Given that $\succ_0$ is drawn from a uniform distribution over all permutations of $C$, we can derive the following inequality for the probability of encountering Type-1-a unsuccessful termination, denoted as $\mathcal{E}_{\ell}$, for a particular length $\ell$ of the rejection chain:
\begin{align*}
    \mathrm{Pr}\bigl[\mathcal{E}_{\ell}^{\mathrm{a}}\bigr]
    &\leq \sum_{\succ_0 \in S'}\mathrm{Pr}\bigl[\mathcal{E}_{\ell}^{\mathrm{a}} \mid \succ_0 \bigr]\cdot\mathrm{Pr}[\succ_0] \\
    &\leq \frac{16\sigma \phi^2}{m}\sum_{\succ_0 \in S'}\mathrm{Pr}[\succ_0] \\
    &= \frac{16\sigma \phi^2}{m},
\end{align*}
where $S'$ denotes all permutations on the set of children $C$ that is used to generate $\succ_0$.

To obtain the overall probability of Type-1-a unsuccessful termination, we sum up the probabilities for all possible lengths $\ell$ and for all children $F^S$. Recall that the length of each child's preference ordering is bounded by $L$, and the length of a rejection chain is upper bounded by $(1-\alpha)n\cdot L$ and lower bounded by $3$. 
Thus, the probability that there exists a rejection cycle leading Type-1-a unsuccessful termination is bounded from above by 
\begin{align*}
    \alpha n\cdot\sum_{\ell=3}^{(1-\alpha)n L}\mathrm{Pr}\bigl[\mathcal{E}_{\ell}^{\mathrm{a}} \bigr]
    \leq 16\alpha(1-\alpha)L \sigma \frac{n^2 \phi^2}{m}.
\end{align*}

If $\phi=O(\log n/n)$, the probability of there being a Type-1-a unsuccessful termination is $ O\left(\frac{(\log n)^2}{n}\right)$, which converges to $0$ as $n$ approaches infinity.
\end{proof}

\subsection{Proof of Lemma~\ref{lemma:type-1-b}}\label{appendix:proof_of_type_1_b}

\LemmaTypeOneB*

\begin{proof}
We next proceed to Type-1-b unsuccessful termination, where a rejection chain is denoted as $c_1 \rightarrow c_2^* \rightarrow \cdots \rightarrow c_{\ell}^* \rightarrow c_1'$. Here, $c_1$ and $c_1'$ are siblings of the same family $f\in F^S$, while $c_2^*, c_3^*, \ldots, c_{\ell}^*$ are children without siblings. Suppose that $c_i^*$ applies to $d_i^*$ for each $i=2,3,\ldots,\ell-1$.

If children $c_1$ and $c'_1$ have nearly identical priorities in $\succ_0$ ($\mathrm{diam}_f \leq |C_f|$), the analysis aligns with that of Type-1-a unsuccessful termination. Consequently, in this scenario, the probability of the rejection chain occurring is at most $16\sigma \phi^2 / m$ for any $\succ_0$ and for any $2 \leq \ell \leq (1-\alpha)n L$.

If children $c_1$ and $c'_1$ have significantly different priorities in $\succ_0$ ($\mathrm{diam}f > |C_f|$), then it only occurs with a probability at most $1 / n^{1+\varepsilon}$ ($\varepsilon>0$). Therefore, even in the worst-case scenario where $\succ_0$ satisfies $c_{1}^* \succ_0 c_{2}^* \succ_0\cdots \succ_0 c_{\ell}^* \succ_0 c_1'^*$, the probability that the last child $c_\ell^*$ causes $c'_1$ to be rejected, is bounded by $\frac{\sigma}{n^{1+\varepsilon}m}$.

Let $\mathcal{E}_{\ell}^{\mathrm{b}}$ denote the event where the rejection chain of length $\ell$ starting with $c_1$ and ending with $c_1'$ occurs. For any $\ell$ and $\succ_0$, we have 
\[\mathrm{Pr}\bigl[\mathcal{E}_{\ell}^{\mathrm{b}} \mid \succ_0 \bigr] \leq \frac{16\sigma \phi^2}{m} + \frac{\sigma}{n^{1+\varepsilon}m}.
\]
We next sum up the probabilities for all possible lengths $\ell$ and for any two children in families with multiple children. The probability of Type-1-b unsuccessful termination occurring is bounded by
\begin{align*}
    &\alpha n \cdot \binom{\bar{k}}{2}\cdot\sum_{\ell=2}^{(1-\alpha)n L}\mathrm{Pr}\bigl[\mathcal{E}_{\ell}^{\mathrm{b}}\bigr] \\
    &\quad\leq \alpha(1-\alpha)L\bar{k}^2 n^2  \left(\frac{16\sigma \phi^2}{m} + \frac{\sigma }{n^{1+\varepsilon}m}\right)\\
    &\quad= O\left(\frac{(\log n)^2}{n}\right) + O(n^{-\varepsilon}). 
\end{align*}
Here, we used $m=\Omega(n)$ and $\phi = O(\log n/n)$. 
This concludes that Type-1 unsuccessful termination does not happen with high probability.
\qedhere
\end{proof}

\subsection{Proof of Lemma~\ref{lem:type-2-lem}}\label{appendix:proof_lemma_type_2_lem}

\LemmaTypeTwoLemma*

\begin{proof}
Consider any two families $f, f' \in F^S$ that do not nest with each other. Without loss of generality, we assume that $f$ top-dominates $f'$, and $f'$ does not dominate $f$, otherwise they would nest with each other. 
Then we have,  
\begin{equation}
\label{formula:lemma}
   \forall c\in C_f, \forall c' \in C(f'), c \succ_0 c'. 
\end{equation}
Suppose $f'$ appears before $f$ in the order $\pi$ over families $F^S$, and $f'$ is currently matched. When $f$ is inserted into the market, we observe that the probability of $f$ causing the rejection of $f'$ is bounded by $\sigma/m$, i.e., $\mathrm{Pr}\bigl[\text{$f$ rejects $f'$}\bigr] \leq \sigma/m$, given that preferences are uniformly bounded.

Next, consider a new order $\pi'$ in which $f$ is placed before $f'$. We aim to analyze the probability of $f'$ causing the rejection of $f$ in a rejection chain of length $\ell$.

We begin with $\ell=2$.
Suppose a child $c \in C_f$ is currently matched to daycare $d_1$, and another child $c' \in C(f')$ also applies to daycare $d_1$, resulting in the rejection of child $c$. As shown in Formula~\eqref{formula:lemma}, we have $c \succ_0 c'$. Since $c' \succ_{1} c$, we can deduce that $\mathrm{Pr}[c' \succ_1 c\mid \succ_0]\leq 4\phi$ from Lemma~\ref{lem:levy}.

Let $\mathcal{E}_0'$ be the event where $f$ rejects $f'$, followed by $f'$ rejecting $f$. The probability that one child in $C(f')$ applies to $d_1$ is upper-bounded by $\sigma/m$. Therefore, we can derive:
    \[
        \mathrm{Pr}\bigl[\mathcal{E}_0'\bigr] \leq \left(\frac{\sigma}{m}\right)^2 4\phi = \frac{4\sigma^2 \phi}{m^2}.
    \]

Next, we consider the scenario where a rejection chain of length $\ell+2$ occurs, where $\ell$ represents the number of children without siblings participating in the rejection chain. Suppose the rejection chain follows the pattern $c \rightarrow c^*_1 \rightarrow c^*_2 \rightarrow \cdots \rightarrow c^*_{\ell} \rightarrow c'$, where $c^*_1,c^*_2,\ldots,c^*_{\ell} \in C^O$. In this case, we have $1 \leq \ell \leq (1-\alpha)n L$.

Let $\mathcal{E}_{\ell}'$ be the event where $f$ rejects $f'$, and subsequently $f'$ rejects $f$ using a rejection chain of length $\ell$.
For any $\succ_0$, the replacement by the Mallows distribution must happen at least twice. Thus, for each $\ell = 1,2,\ldots,(1-\alpha)n L$, we have
\[
    \mathrm{Pr}\bigl[\mathcal{E}_{\ell}' \mid \succ_0 \bigr] \leq \left(\frac{\sigma' }{m}\right)^2 16\phi^2 \leq \frac{16\sigma' \phi^2}{m^2}.
\]
We sum up the probabilities for all possible $\succ_0$, and achieve $\mathrm{Pr}\bigl[\mathcal{E}_{\ell}' \bigr] \leq \frac{16\sigma' \phi^2}{m^2}$ for each $\ell = 1,2,\ldots,(1-\alpha)n L$. Then we obtain
    \[
        \sum_{\ell=1}^{(1-\alpha)n L}\mathrm{Pr}\bigl[\mathcal{E}_{\ell}'\bigr] \leq \frac{16(1-\alpha)L\sigma n \phi^2}{m^2}.
    \]
Finally, since $m=\Omega(n)$ and $\phi = O(\log n / n)$, we get
    \begin{align*}
        &\mathrm{Pr}\bigl[\text{there exists a pair of families with siblings} \\
        &\quad \text{that cause rejections with each other}\bigr] \\
        &\quad= \sum_{f,f'\in F^S}\mathrm{Pr}\left[\bigcup_{\ell=0}^{(1-\alpha)\bar{k}n}\mathcal{E}_{\ell}'\right] \\
        &\quad \leq \sum_{f,f'\in F^S}\sum_{\ell=0}^{(1-\alpha)n L}\mathrm{Pr}\bigl[\mathcal{E}_{\ell}'\bigr] \\
        &\quad = \sum_{f,f'\in F^S}\left(\mathrm{Pr}\bigl[\mathcal{E}_{0}'\bigr] + \sum_{\ell=1}^{(1-\alpha)n L}\mathrm{Pr}\bigl[\mathcal{E}_{\ell}'\bigr] \right)\\
        &\quad \leq (\alpha n)^2\left(\frac{16\sigma \phi}{m^2} + \frac{16(1-\alpha)\bar{k}\sigma n \phi^2}{m^2} \right) \\
        &\quad = O\left(\frac{\log n}{n}\right).\qedhere
    \end{align*}
\end{proof}

\subsection{Proof of Lemma~\ref{lemma:type-2}}

\LemmaTypeTwo*

\begin{proof}
We first consider the probability that any two pairs of families with multiple siblings nest with each other w.r.t. the reference ordering $\succ_0$.

For any two families $f$ and $f'$, if they nest with each other, then the diameters of both $f$ and $f'$ are large, i.e., $\mathrm{diam}_f > |C_f|$ and $\mathrm{diam}_{f'} > |C(f')|$. Thus, 
the inequality $\mathrm{Pr}\bigl[\mathrm{diam}_f \geq |C_f|\bigr] \leq \frac{1}{n^{1+\varepsilon}}$ implies that
\[
    \mathrm{Pr}\bigl[\text{$f$ and $f'$ nest with each other}\bigr] 
    \leq \left(\frac{1}{n^{1+\varepsilon}}\right)^{2}.
\]
Hence, we have
\begin{align*}
    &\mathrm{Pr}\bigl[\text{there exist two families who nest with each other}\bigr] \\
    &\quad\leq \sum_{f,f'\in F^S}\mathrm{Pr}\bigl[\text{$f$ and $f'$ nest with each other}\bigr] \\
    &\quad\leq\binom{\alpha n}{2} \cdot \left(\frac{1}{n^{1+\varepsilon}}\right)^{2}  \\
    &\quad\leq \alpha^2 n^2 \cdot \left(\frac{1}{n^{1+\varepsilon}}\right)^{2}  \\
    &\quad= O\bigl(n^{-2\varepsilon}\bigr).
\end{align*}
Since $\varepsilon>0$ is a constant, the probability that any two families do not nest with each other approaches $1$ as $n$ tends to infinity.

We now upper-bound the probability of Type-2 unsuccessful termination. In cases where two families nest with each other, Type-2 unsuccessful termination may occur with a constant probability. However, we have demonstrated that the probability of two families nesting with each other is at most $O(n^{-2\varepsilon})$. In instances where no two families nest with each other, Type-2 unsuccessful termination happens with a probability of at most $O(\log n /n)$ as shown in Lemma~\ref{lem:type-2-lem}. Therefore, we can express the probability of Type-2 unsuccessful termination as follows:
\begin{align*}
    \mathrm{Pr}\bigl[\text{Type-2 unsuccessful termination happens}\bigr] =O\bigl(n^{-2\varepsilon}\bigr) + O(\log n /n).
\end{align*}
This completes the proof.
\end{proof}

Lemma~\ref{lemma:type-1-a},~\ref{lemma:type-1-b} and~\ref{lemma:type-2} imply the existence of a stable matching with high probability for the large random market, thus concluding the proof of Theorem~\ref{theorem:main}. 

\section{Experiments}
In this section, we conduct comprehensive experiments to address three key questions: (1) How often does a stable matching exist in a large random market?  
(2) How effective is our proposed ESDA algorithm in identifying stable matchings? 
(3) How does our stronger stability concept affect the existence of stable matchings compared to ABH-stability?  

Given the limitations of the ESDA algorithm in computing stable matchings in certain scenarios, we adopt a constraint programming (CP) approach as an alternative. This method reliably produces a stable matching whenever one exists \citep{SYT+24a}. We compare our algorithm against the original SDA, CP with ABH stability, and CP with our proposed stability concept. 
To evaluate these algorithms, we use both real-world and synthetic datasets, focusing on two key aspects: the frequency with which each algorithm identifies a stable matching and their running time. All algorithms are implemented in Python and executed on a standard laptop without additional computational resources.

The experimental findings can be summarized as follows:  
(1) As established in Theorem 1, a stable matching is highly likely to exist when daycares share similar priority orderings over children.  
(2) The ESDA algorithm achieves performance close to the optimal solution, without experiencing a significant performance decline compared to SDA, while satisfying a stronger stability concept.  
(3) In general, our proposed stability concept does not reduce the probability of stable matchings existing compared to the ABH stability concept.

\subsection{Experiments on Real-life Datasets}
\label{appendix:real_data}

We first evaluate our algorithm on six real-world datasets obtained from three municipalities. All four algorithms including SDA, ESDA, CP-ABH, and CP-ours, successfully identify a stable matching in these cases.

We are collaborating with several municipalities in Japan, and as part of our collaboration, we provide a detailed description of the practical daycare matching markets based on data sets provided by three representative municipalities.

%

\paragraph{The Number of Children}
The number of children in each market varies from 500 to 1600, with the proportion of children having siblings consistently spanning from 15$\%$ to 20$\%$, as shown in Table~\ref{table:frac_siblings}.
\begin{table}[H]
\begin{center}
\begin{tabular}{c|rr}
    \hline
    Dataset & Siblings  & $\#$Children \\
    \hline
    Shibuya-21 & 16.24$\%$ & 1589
    \\
    \hline
    Shibuya-22 & 15.38$\%$ & 1372 
    \\
    \hline
    Tama-21 &  16.45$\%$ & 635
    \\
    \hline
    Tama-22 &  16.00$\%$  & 550
    \\
    \hline
    Koriyama-22 & 20.68$\%$ & 1383
    \\
    \hline 
    Koriyama-23 & 19.14$\%$ & 1458
    \\
    \hline    
\end{tabular}
\end{center}
\caption{Dataset statistics showing the percentage of children with siblings and total number of children. }
\label{table:frac_siblings}
\end{table}

\paragraph{Length of Preferences}
The preference ordering of an only child is relatively short compared to the available daycares, averaging between 3 and 4.5 choices. Similarly, children from families with siblings exhibit an average of 3 to 4.5 distinct daycares in their individual preferences. Furthermore, siblings within the same family often share a similar set of daycares in their joint preference ordering. The details are presented in Table~\ref{table:ave_pref_len}.
\begin{table}[H]
\begin{center}
\begin{tabular}{c|rrrr}
    \hline
    Dataset & Single & Siblings & Distinct & $\#$Daycares \\
    \hline
    Shibuya-21 & 4.45 & 14.86 & 4.26 & 72
    \\
    \hline
    Shibuya-22 & 3.76 & 6.58 & 3.64 & 72
    \\
    \hline
    Tama-21 & 3.29 & 38.29 & 3.43 & 33
    \\
    \hline
    Tama-22 & 3.01 & 8.55 & 3.17 & 33 
    \\
    \hline
    Koriyama-22 & 3.02 & 21.38 & 3.60 & 86
    \\
    \hline 
    Koriyama-23 & 3.10 & 9.42 & 3.13 & 86
    \\
    \hline    
\end{tabular}
\end{center}
\caption{Preference list statistics across datasets. Single and Siblings columns show average preference list lengths for families with one child and multiple children respectively. Distinct shows the average number of unique daycares in sibling preference lists. $\#$Daycares shows the total number of daycares in each dataset.}
\label{table:ave_pref_len}
\end{table}

\paragraph{The Number of Families with Siblings and Twins}
A critical aspect not addressed in Section~\ref{sec:model} is that each child is associated with an age ranging from 0 to 5. Inspired by prior work \citep{STM+23a,SYT+24a}, we assume that there are six copies of the same daycare, each catering to a specific age group. The distribution of children in the market is uneven, with a significant majority aged 0 and 1. In Table~\ref{table:twins}, we present the count of families with siblings, including twins (i.e., pairs of siblings of the same age).

\begin{table}[H]
\begin{center}
\begin{tabular}{c|rr|rr}
    \hline
    & \multicolumn{4}{c}{$\#$Families}\\
    \cmidrule{2-5}
    Dataset & \multicolumn{2}{c|}{2 Children} 
    & \multicolumn{2}{c}{$\ge$3 Children} \\
    \cmidrule{2-5}
     & Total & Twins & Total & Twins \\
    \hline
    Shibuya-21    & 120 & 14 & 6 & 4 \\
    Shibuya-22    & 101 & 25 & 3 & 3 \\
    Tama-21       & 42  & 3  & 3 & 3 \\
    Tama-22       & 44  & 8  & 0 & 0 \\
    Koriyama-22   & 123 & 10 & 13 & 2 \\
    Koriyama-23   & 130 & 12 & 6  & 0 \\
    \hline    
\end{tabular}
\end{center}
\caption{Number of families with siblings and twins. The second and third columns represent families with two children, while the last two columns represent families with three or more children.}
\label{table:twins}
\end{table}

\paragraph{Demand-Supply Imbalance Across Age Groups}
Despite the total capacity of all daycares exceeding the number of applicants, there is a significant imbalance between demand and supply across different age groups. Specifically, there is a shortage of slots for children aged 0, 1 and 2, while there is a surplus of slots for children aged 4 and 5, as shown in Table~\ref{table:demand}.
\begin{table}[H]
\begin{center}
\begin{tabular}{c|c|rrrrrr}
\hline
\multirow{2}{*}{Dataset} & \multirow{2}{*}{Type} & \multicolumn{6}{c}{Age} \\
\cmidrule{3-8}
& & 0 & 1 & 2 & 3 & 4 & 5 \\
\hline
\multirow{2}{*}{Shibuya-21} 
& Applicants & 569 & 656 & 171 & 136 & 37 & 20 \\
& Capacity   & 509 & 613 & 239 & 265 & 268 & 275 \\
\hline
\multirow{2}{*}{Shibuya-22}
& Applicants & 540 & 582 & 134 & 67 & 33 & 16 \\
& Capacity   & 497 & 586 & 186 & 233 & 255 & 306 \\
\hline
\multirow{2}{*}{Tama-21}
& Applicants & 181 & 257 & 98 & 75 & 17 & 7 \\
& Capacity   & 241 & 222 & 123 & 106 & 57 & 68 \\
\hline
\multirow{2}{*}{Tama-22}
& Applicants & 181 & 219 & 91 & 43 & 8 & 8 \\
& Capacity   & 231 & 218 & 100 & 97 & 45 & 47 \\
\hline
\multirow{2}{*}{Koriyama-22}
& Applicants & 379 & 538 & 140 & 231 & 59 & 36 \\
& Capacity   & 546 & 585 & 220 & 327 & 276 & 171 \\
\hline
\multirow{2}{*}{Koriyama-23}
& Applicants & 366 & 588 & 167 & 239 & 64 & 33 \\
& Capacity   & 559 & 511 & 218 & 282 & 139 & 188 \\
\hline
\end{tabular}
\end{center}
\caption{Distribution of applicants and daycare capacity by age across datasets.}
\label{table:demand}
\end{table}

\paragraph{Similar Priority Orderings Across Daycares}
Municipalities assign priority scores to children, with siblings from the same family typically sharing identical scores. Daycares then make slight adjustments to these priority scores to establish a strict priority ordering. As a result, all daycares generally have similar priority orderings for the children.

\subsection{Experiments on Synthetic Datasets}
\label{subsection:experiments:synthetic}

\begin{table}[htb]
\begin{center}
\scalebox{0.9}{
\begin{tabular}{ccrrrrrr}
\hline
$\#$children & Algorithm & Success & Time (s) & Success & Time (s) & Success & Time (s) \\
\hline
\hline
& & \multicolumn{2}{c}{$\phi = 0.0$} & \multicolumn{2}{c}{$\phi = 0.3$} & \multicolumn{2}{c}{$\phi = 0.5$} \\
\cmidrule{2-8}
\cmidrule(lr){3-4} \cmidrule(lr){5-6} \cmidrule(lr){7-8}
\cmidrule{2-8}
500& SC & 0/100 & nan $\pm$ nan & 0/100 & nan $\pm$ nan & 0/100 & nan $\pm$ nan \\
& SDA & 99/100 & 0.04 $\pm$ 0.01 & 100/100 & 0.04 $\pm$ 0.01 & 100/100 & 0.04 $\pm$ 0.01 \\
& ESDA & 99/100 & 0.04 $\pm$ 0.01 & 100/100 & 0.04 $\pm$ 0.01 & 99/100 & 0.04 $\pm$ 0.01 \\
& CP-ABH & 100/100 & 0.72 $\pm$ 0.02 & 100/100 & 0.75 $\pm$ 0.02 & 100/100 & 0.77 $\pm$ 0.02 \\
& CP-Ours & 100/100 & 0.72 $\pm$ 0.02 & 100/100 & 0.75 $\pm$ 0.02 & 100/100 & 0.77 $\pm$ 0.02 \\
\cmidrule{2-8}
& & \multicolumn{2}{c}{$\phi = 0.7$} & \multicolumn{2}{c}{$\phi = 0.9$} & \multicolumn{2}{c}{$\phi = 1.0$} \\
\cmidrule(lr){3-4} \cmidrule(lr){5-6} \cmidrule(lr){7-8}
\cmidrule{2-8}
& SC & 0/100 & nan $\pm$ nan & 0/100 & nan $\pm$ nan & 0/100 & nan $\pm$ nan \\
& SDA & 100/100 & 0.04 $\pm$ 0.01 & 98/100 & 0.04 $\pm$ 0.01 & 80/100 & 0.04 $\pm$ 0.01 \\
& ESDA & 100/100 & 0.04 $\pm$ 0.01 & 97/100 & 0.04 $\pm$ 0.01 & 76/100 & 0.05 $\pm$ 0.01 \\
& CP-ABH & 100/100 & 0.72 $\pm$ 0.02 & 99/100 & 0.66 $\pm$ 0.02 & 98/100 & 0.71 $\pm$ 0.02 \\
& CP-Ours & 100/100 & 0.72 $\pm$ 0.02 & 99/100 & 0.68 $\pm$ 0.02 & 94/100 & 0.72 $\pm$ 0.02 \\
\hline
\hline
& & \multicolumn{2}{c}{$\phi = 0.0$} & \multicolumn{2}{c}{$\phi = 0.3$} & \multicolumn{2}{c}{$\phi = 0.5$} \\
\cmidrule(lr){3-4} \cmidrule(lr){5-6} \cmidrule(lr){7-8}
\cmidrule{2-8}
1000& SC & 0/100 & nan $\pm$ nan & 0/100 & nan $\pm$ nan & 0/100 & nan $\pm$ nan\\
& SDA & 100/100 & 0.16 $\pm$ 0.03 & 100/100 & 0.16 $\pm$ 0.03 & 100/100 & 0.16 $\pm$ 0.02 \\
& ESDA & 100/100 & 0.17 $\pm$ 0.03 & 100/100 & 0.17 $\pm$ 0.03 & 100/100 & 0.17 $\pm$ 0.03 \\
& CP-ABH & 100/100 & 1.92 $\pm$ 0.04 & 100/100 & 1.95 $\pm$ 0.04 & 100/100 & 1.96 $\pm$ 0.04  \\
& CP-Ours & 100/100 & 1.93 $\pm$ 0.03 & 100/100 & 1.95 $\pm$ 0.04 & 100/100 & 1.99 $\pm$ 0.05 \\
\hline
\cmidrule{2-8}
& & \multicolumn{2}{c}{$\phi = 0.7$} & \multicolumn{2}{c}{$\phi = 0.9$} & \multicolumn{2}{c}{$\phi = 1.0$} \\
\cmidrule(lr){3-4} \cmidrule(lr){5-6} \cmidrule(lr){7-8}
\cmidrule{2-8}
& SC & 0/100 & nan $\pm$ nan & 0/100 & nan $\pm$ nan & 0/100 & nan $\pm$ nan \\
& SDA & 100/100 & 0.15 $\pm$ 0.02 & 97/100 & 0.15 $\pm$ 0.03 & 77/100 & 0.18 $\pm$ 0.03 \\
& ESDA & 100/100 & 0.16 $\pm$ 0.03 & 97/100 & 0.16 $\pm$ 0.03 & 73/100 & 0.20 $\pm$ 0.04 \\
& CP-ABH & 100/100 & 1.89 $\pm$ 0.03 & 99/100 & 1.90 $\pm$ 0.04 & 97/100 & 1.98 $\pm$ 0.03 \\
& CP-Ours & 100/100 & 1.93 $\pm$ 0.04 & 99/100 & 1.91 $\pm$ 0.04 & 91/100 & 1.98 $\pm$ 0.03 \\
\hline
\hline
& & \multicolumn{2}{c}{$\phi = 0.0$} & \multicolumn{2}{c}{$\phi = 0.3$} & \multicolumn{2}{c}{$\phi = 0.5$} \\
\cmidrule(lr){3-4} \cmidrule(lr){5-6} \cmidrule(lr){7-8}
\cmidrule{2-8}
3000& SC & 0/100 & nan $\pm$ nan & 0/100 & nan $\pm$ nan & 0/100 & nan $\pm$ nan\\
& SDA & 100/100 & 1.39 $\pm$ 0.14 & 100/100 & 1.39 $\pm$ 0.13 & 100/100 & 1.40 $\pm$ 0.12 \\
& ESDA & 100/100 & 1.48 $\pm$ 0.16 & 100/100 & 1.48 $\pm$ 0.16 & 100/100 & 1.48 $\pm$ 0.14 \\
& CP-ABH & 100/100 & 11.72 $\pm$ 0.13 & 100/100 & 11.84 $\pm$ 0.14 & 100/100 & 11.81 $\pm$ 0.13  \\
& CP-Ours & 100/100 & 11.77 $\pm$ 0.13 & 100/100 & 11.82 $\pm$ 0.14 & 100/100 & 11.90 $\pm$ 0.15  \\
\cmidrule{2-8}
& & \multicolumn{2}{c}{$\phi = 0.7$} & \multicolumn{2}{c}{$\phi = 0.9$} & \multicolumn{2}{c}{$\phi = 1.0$} \\
\cmidrule(lr){3-4} \cmidrule(lr){5-6} \cmidrule(lr){7-8}
\cmidrule{2-8}
& SC & 0/100 & nan $\pm$ nan & 0/100 & nan $\pm$ nan & 0/100 & nan $\pm$ nan \\
& SDA & 100/100 & 1.41 $\pm$ 0.15 & 99/100 & 1.42 $\pm$ 0.13 & 74/100 & 1.69 $\pm$ 0.22 \\
& ESDA & 100/100 & 1.50 $\pm$ 0.17 & 99/100 & 1.51 $\pm$ 0.15 & 73/100 & 1.87 $\pm$ 0.25 \\
& CP-ABH & 100/100 & 12.00 $\pm$ 0.14 & 100/100 & 11.58 $\pm$ 0.14 & 100/100 & 11.86 $\pm$ 0.11 \\
& CP-Ours & 100/100 & 11.97 $\pm$ 0.12 & 100/100 & 11.64 $\pm$ 0.14 & 95/100 & 11.86 $\pm$ 0.15 \\
\hline
\hline
\end{tabular}
}
\end{center}
\caption{
Performance comparison across different children sizes ($|C|=500,1000,3000$) and dispersion parameters ($\phi$) in the Mallows model.
SC is the Sequential Couples algorithm~\citep{KPR13a}.
SDA is the Sorted Deferred Acceptance algorithm introduced by \citet{ABH14a}.
ESDA (Extended SDA) is our proposed extension of the SDA algorithm.
Both CP-ABH and CP-Ours use constraint programming to find stable matchings where CP-ABH uses ABH-stability from \citep{ABH14a} and CP-Ours uses our proposed notion of stability as constraints.
Success shows the number of successful runs out of 100 instances.
Time shows mean,$\pm$,std computation time in seconds for successful runs only.}
\label{table:synthetic_results}
\end{table}

\begin{table}[htb]
\begin{center}
\scalebox{0.9}{
\begin{tabular}{ccrrrrrr}
\hline
$\#$children & Algorithm & Success & Time (s) & Success & Time (s) & Success & Time (s) \\
\hline
\hline
& & \multicolumn{2}{c}{$\phi = 0.0$} & \multicolumn{2}{c}{$\phi = 0.3$} & \multicolumn{2}{c}{$\phi = 0.5$} \\
\cmidrule(lr){3-4} \cmidrule(lr){5-6} \cmidrule(lr){7-8}
\cmidrule{2-8}
5000& SC & 0/100 & nan $\pm$ nan & 0/100 & nan $\pm$ nan & 0/100 & nan $\pm$ nan\\
& SDA & 100/100 & 3.89 $\pm$ 0.30 & 100/100 & 3.91 $\pm$ 0.28 & 100/100 & 3.86 $\pm$ 0.29 \\
& ESDA & 100/100 & 4.13 $\pm$ 0.35 & 100/100 & 4.15 $\pm$ 0.31 & 100/100 & 4.11 $\pm$ 0.34 \\
& CP-ABH & 100/100 & 29.14 $\pm$ 0.32 & 100/100 & 29.42 $\pm$ 0.32 & 100/100 & 29.46 $\pm$ 0.30  \\
& CP-Ours & 100/100 & 29.24 $\pm$ 0.27 & 100/100 & 29.33 $\pm$ 0.27 & 100/100 & 29.39 $\pm$ 0.52 \\
\cmidrule{2-8}
& & \multicolumn{2}{c}{$\phi = 0.7$} & \multicolumn{2}{c}{$\phi = 0.9$} & \multicolumn{2}{c}{$\phi = 1.0$} \\
\cmidrule(lr){3-4} \cmidrule(lr){5-6} \cmidrule(lr){7-8}
\cmidrule{2-8}
& SC & 0/100 & nan $\pm$ nan & 0/100 & nan $\pm$ nan & 0/100 & nan $\pm$ nan \\
& SDA & 100/100 & 3.93 $\pm$ 0.32 & 100/100 & 3.88 $\pm$ 0.29 & 76/100 & 4.68 $\pm$ 0.32 \\
& ESDA & 99/100 & 4.17 $\pm$ 0.36 & 100/100 & 4.11 $\pm$ 0.32 & 74/100 & 5.17 $\pm$ 0.39 \\
& CP-ABH & 100/100 & 28.55 $\pm$ 0.30 & 100/100 & 29.04 $\pm$ 0.31 & 95/100 & 30.49 $\pm$ 6.37 \\
& CP-Ours & 100/100 & 28.61 $\pm$ 0.36 & 100/100 & 31.09 $\pm$ 20.27 & 93/100 & 29.64 $\pm$ 0.35 \\
\hline
\hline
& & \multicolumn{2}{c}{$\phi = 0.0$} & \multicolumn{2}{c}{$\phi = 0.3$} & \multicolumn{2}{c}{$\phi = 0.5$} \\
\cmidrule(lr){3-4} \cmidrule(lr){5-6} \cmidrule(lr){7-8}
\cmidrule{2-8}
10000& SC & 0/100 & nan $\pm$ nan & 0/100 & nan $\pm$ nan & 0/100 & nan $\pm$ nan\\
& SDA & 100/100 & 16.79 $\pm$ 0.92 & 100/100 & 16.36 $\pm$ 0.89 & 100/100 & 16.35 $\pm$ 0.91 \\
& ESDA & 100/100 & 17.62 $\pm$ 1.11 & 100/100 & 17.57 $\pm$ 0.97 & 100/100 & 17.35 $\pm$ 1.07 \\
& CP-ABH  & 100/100 & 106.69 $\pm$ 1.61 & 100/100 & 103.02 $\pm$ 1.53 & 100/100 & 104.83 $\pm$ 0.90 \\
& CP-Ours & 100/100 & 105.84 $\pm$ 0.88 & 100/100 & 104.69 $\pm$ 0.88 &  100/100 & 104.60 $\pm$ 0.75 \\
\cmidrule{2-8}
& & \multicolumn{2}{c}{$\phi = 0.7$} & \multicolumn{2}{c}{$\phi = 0.9$} & \multicolumn{2}{c}{$\phi = 1.0$} \\
\cmidrule(lr){3-4} \cmidrule(lr){5-6} \cmidrule(lr){7-8}
\cmidrule{2-8}
& SC & 0/100 & nan $\pm$ nan & 0/100 & nan $\pm$ nan & 0/100 & nan $\pm$ nan \\
& SDA & 100/100 & 16.61 $\pm$ 0.94 & 100/100 & 16.32 $\pm$ 0.89 & 68/100 & 19.65 $\pm$ 1.23 \\
& ESDA & 100/100 & 17.57 $\pm$ 1.03 & 100/100 & 17.30 $\pm$ 1.05 & 66/100 & 21.63 $\pm$ 1.49 \\
& CP-ABH & 100/100 & 103.83 $\pm$ 0.97 & 100/100 & 104.95 $\pm$ 0.73 & 95/100 & 105.01 $\pm$ 1.11\\
& CP-Ours & 100/100 & 105.72 $\pm$ 1.05 & 100/100 & 105.30 $\pm$ 0.70 & 85/100 & 104.55 $\pm$ 0.64 \\
\hline
\hline
\end{tabular}
}
\end{center}
\caption{Performance comparison across different children sizes ($|C|=5000,10000$) and dispersion parameters ($\phi$) in the Mallows model. 
}
\label{table:synthetic_results_5000}
\end{table}

We outline the steps to generate synthetic datasets.

We define the total number of children, denoted by $|C|$, drawn from the set $\{500, 1000, 3000, 5000, 10000\}$. 
We assume that the proportion of children with siblings is bounded by $\alpha = 0.2$.  
For families with siblings, we consider only two-sibling families and three-sibling families, where the children account for $80\%$ and $20\%$, respectively.  

The number of two-sibling families is calculated as  
\[
|F^S_2| = \text{int}\left(\frac{\alpha \times |C| \times 0.8}{2}\right),
\]  
and the number of three-sibling families is calculated as  
\[
|F^S_3| = \text{int}\left(\frac{\alpha \times |C| \times 0.2}{3}\right).
\]  

As a result, approximately $20\%$ of the children belong to families with siblings ($\alpha \approx 0.2$).  

The total number of children with siblings is calculated as  
\[
|C^S| = (|F^S_2| \times 2 + |F^S_3| \times 3),
\]  
while the number of children without siblings is  
\[
|C^O| = |C| - |C^S|.
\]  

Correspondingly, the total number of families is  
\[
|F| = |F^O| + |F^S|,
\]  
where $|F^O|$ and $|F^S|$ represent the numbers of families without and with siblings, respectively.  

The number of daycares is determined as  
\[
|D| = \text{int}(0.1 \times |F|),
\]  
and the capacity of each daycare is fixed using the list  
$[5, 5, 1, 1, 1, 1]$, where each element corresponds to a specific age group in the range from 0 to 5.

For each child without siblings ($C^O$), we randomly assign preferences for 5 daycares from the set $D$. For families with siblings ($F^S$), we generate an individual preference ordering of length 10 for each child $c \in C_f$ by uniformly sampling from $D$. Subsequently, we consider all possible combinations of preferences within the family and uniformly select a joint preference ordering of length 10.  

We vary the dispersion parameter $\phi$ within the range $\{0.0, 0.3, 0.5, 0.7, 0.9, 1.0\}$ while keeping the parameter $\varepsilon$, which is used to generate a reference ordering $\succ_0$, fixed at 1. For each specified setting, we generate 100 instances.  

In addition to the stability analysis, we conducted a comparison of the running times between these algorithms. Although SDA and ESDA may need to check all permutations of $F^S$ in the worst-case scenario, it consistently demonstrated significantly faster performance than the CP algorithm across all cases.

Regarding the experimental findings:  
(i) A stable matching is very likely to exist for $\phi \leq 0.9$ and with high probability for $\phi = 1.0$.  
(ii) For $\phi \leq 0.9$, the ESDA algorithm consistently identified a stable matching, while the SDA algorithm consistently identified an ABH-stable matching.  
(iii) Although our stability concept is stronger than ABH-stability, there is no significant decrease in the existence of stable matchings for most of the settings, except for the case where $|C| = 10000$ and $\phi = 1.0$, which is unlikely to occur in practice.

We summarize the experimental results in Table~\ref{table:synthetic_results} and Table~\ref{table:synthetic_results_5000}.

\section{Conclusion}

In this study, we investigate the reasons behind the existence of stable matchings in practical daycare markets, identifying the shared priority ordering among daycares as a primary factor. Our contributions include a probabilistic analysis of such large random markets and the introduction of the ESDA algorithm to identify stable matchings. Experimental results demonstrate the efficiency of the ESDA algorithm under various conditions. We plan to continue this study by investigating additional factors that contribute to the existence of stable matchings in more general settings, beyond the case of similar priority orderings over children.

\section*{Acknowledgments}
This work was partially supported by the Japan Science and Technology Agency under the ERATO Grant Number JPMJER2301.
We thank the anonymous reviewers of NeurIPS 2024 and AAMAS 2025 for their valuable comments.

\bibliographystyle{plainnat}
\bibliography{existence}

\appendix

\end{document}